\documentclass[manuscript]{aastex}	
\usepackage{epsfig}			
\usepackage{graphicx,color}		
\usepackage{amssymb}			
\usepackage{color}			
\usepackage{url}			
\usepackage{amsmath}			
\usepackage{rotating}			
\usepackage{float}			
\usepackage{textcomp}			
\usepackage{psfig}
\usepackage{epstopdf}
\usepackage{dcolumn}
\usepackage{times}
\usepackage{tabularx}
\usepackage[english]{babel}
\usepackage[normalem]{ulem}
\usepackage{float}
\usepackage{ragged2e}

\shorttitle{Reflection of Slow Magneto-acoustic wave in hot coronal loop}
\shortauthors{S. Mandal et al.}

\begin{document}

\title{Reflection Of Propagating Slow Magneto-acoustic Waves In Hot Coronal Loops : Multi-instrument Observations and Numerical Modelling }

\author{Sudip Mandal$^{1}$,
Ding Yuan$^{2,3}$,
Xia Fang$^{2}$,
Dipankar Banerjee$^{1,4}$,
Vaibhav Pant$^{1}$,
Tom Van Doorsselaere$^{2}$}
 
\affil{$^{1}$Indian Institute of Astrophysics, Koramangala, Bangalore 560034, India. e-mail: {\color{blue}{sudip@iiap.res.in}}\\
$^{2}$Centre for mathematical Plasma Astrophysics, Department of Mathematics, KU Leuven, Celestijnenlaan 200B, bus 2400, 3001, Leuven, Belgium. e-mail: {\color{blue}{xia.fang@wis.kuleuven.be}}\\
$^{3}$ Jeremiah Horrocks Institute, University of Central Lancashire, Preston, PR1 2HE ,UK\\
$^{4}$ Center of Excellence in Space Sciences India, IISER Kolkata, Mohanpur 741246, West Bengal, India}

\justify
\begin{abstract}

Slow MHD waves are important tools for understanding the coronal structures and dynamics. In this paper, we report a number of observations, from X-Ray Telescope (XRT) on board HINODE and SDO/AIA of reflecting longitudinal waves in hot coronal loops. To our knowledge, this is the first report of this kind as seen from the XRT and simultaneously with the AIA. The wave appears after a micro-flare occurs at one of the footpoints. We estimate the density and the temperature of the loop plasma by performing DEM analysis on the AIA image sequence. The estimated speed of propagation is comparable or lower than the local sound speed suggesting it to be a propagating slow wave. The intensity perturbation amplitudes, in every case, falls very rapidly as the perturbation moves along the loop and eventually vanishes after one or more reflections. To check the consistency of such reflection signatures with the obtained loop parameters, we perform a 2.5D MHD simulation, which uses the parameters obtained from  our observation as inputs and performed forward modelling to synthesize AIA 94~\r{A} images. Analyzing the synthesized images, we obtain the same properties of the observables as for the real observation. From the analysis we conclude that a footpoint heating can generate slow wave which then reflects back and forth in the coronal loop before fading out. Our analysis on the simulated data shows that the main agent for this damping is the anisotropic thermal conduction.

\end{abstract}


\section{Introduction}

MHD waves play an important role in understanding the solar structures and the coronal heating process \citep{1984ApJ...279..857R,lrsp-2005-3,2007SoPh..246....3B,2012RSPTA.370.3193D}. Extreme ultra-violet (EUV) imaging analysis provides access to the loop diagnostics i.e loop density, temperature, flows including estimation of magnetic field \citep{1984ApJ...279..857R,1999ApJ...520..880A,1999Sci...285..862N,2001A&A...372L..53N,2013A&A...552A.138V}.
Slow MHD waves in the solar corona were first observed by  \citet{1997ApJ...491L.111O,1998ApJ...501L.217D,berghmans1999active} as quasi-periodic propagating disturbances(PDs) channeling through coronal structures. These PDs have an apparent speed close to the sound speed of that medium leading to an explanation of these PDs as the propagating slow waves \citep{2012SoPh..279..427K,2009ApJ...697.1674M,2009ApJ...706L..76M}. Using simultaneous imaging and spectroscopic data, recent analysis, including `Red-Blue (R-B)' asymmetry, shows that the upflow scenario could also explain the observed PDs \citep{2009ApJ...701L...1D,2011ApJ...727L..37T,2015RAA....15.1832M}. Recently it is  shown by \citet{2010ApJ...724L.194V} that the slow waves can also naturally explain the behaviour of R-B parameter.

Standing slow waves have also been studied very closely by many authors. \citet{2002ApJ...574L.101W,2003A&A...406.1105W,2003A&A...402L..17W} observed damped Doppler shift oscillations in hot coronal lines with SoHO/SUMER. They interpreted these oscillatory Doppler shifts of the high temperature ($\mathrm{T} >$ 6 $\mathrm{MK}$ ) Fe \textsc{xix} and Fe \textsc{xxi} lines as standing slow waves generated by an impulsive trigger at one of the loop footpoints.
\citet{2005A&A...435..753W} analyzed 54 Doppler shift oscillations in 27 flare-like events from SOHO/SUMER and interpreted them as being caused by standing slow waves because they exhibit a quarter period phase shift between the intensity and velocity oscillations.~\citet{2005ApJ...620L..67M} reported Doppler shift oscillations in S \textsc{xv} and~Ca \textsc{xix} lines observed with BCS/Yohkoh where they found oscillations with a period of few minutes \citep{2006ApJ...639..484M} (from HINODE/EIS \citet{2008ApJ...681L..41M}). Using forward modelling, \citet{2015ApJ...807...98Y} synthesized SDO/AIA and SoHO/SUMER emissions to study the standing slow wave modes in a hot flaring loop. Apart from recovering the quarter period phase shift between intensity and Doppler velocity, these authors also found asymmetric emission intensity during the positive and negative temperature perturbation phase.

\citet{2005A&A...436..701S} generated slow waves numerically by applying temperature perturbation pulses at different positions in the loop and observed that the generated mode depends upon the location of the pulses. The study of various aspects of the slow wave generation have been extensively carried out using 2D and 3D MHD models \citep{2007ApJ...668L..83S,2009AnGeo..27.3899S,2009IAUS..257..151O,2012ApJ...754..111O}. Using a 3D MHD model \citet{2012ApJ...754..111O} concluded that the impulsive injection of the energy at the active region loop footpoint can excite slow and fast waves simultaneously along with the observed outflows. Damping of the Doppler oscillations (observed with SUMER) have also been studied thoroughly by \citet{2002ApJ...580L..85O}. Using a 1D MHD code simulation, these authors concluded that the thermal conduction plays the significant role in damping of these waves rather than the compressive viscosity. Damping of the waves have been further studied by \citet{2003A&A...408..755D,2004A&A...415..705D,2004A&A...425..741D} where they have included the gravitational stratification, field line divergence and mode coupling apart from thermal conduction and viscosity to damp these waves.

Reflection of slow MHD waves propagating through the coronal loops have been reported recently by \citet{2013ApJ...779L...7K} using high resolution imaging data from the SDO/AIA high temperature 131~\r{A} and 94~\r{A} channels. Slow wave generated at one of the footpoints, was reflected back and forth, a couple of times from the loop footpoints before fading out. The propagation speed of the wave was about 460-510 km s$^{-1}$, which is very close to the sound speed at the temperature obtained from the DEM analysis. \citet{2015arXiv150904536F}, using a 2.5D MHD simulation and forward modelling, reproduced such slow wave reflections in a hot coronal loop. A flare-like instantaneous energy perturbation at the footpoint evaporates a plasma blob which then propagates as a slow wave front. The wave then bounces back and forth along the loop as observed by \citet{2013ApJ...779L...7K}. \citet{2015arXiv150904536F} also used line parameters of the synthesized SUMER Fe line to show that these are propagating mode. With the use of a particle tracer, these authors confirm that such propagating disturbances better agree with a dominant wave scenario along with a mass flow component. Recently \citet{2015ApJ...804....4K} reported quasi-periodic intensity oscillations in AIA extreme ultraviolet (EUV) channels along with the X-ray channel of Fermi gamma ray burst (GRB) monitor. These authors also propose the repetitive reconnection scenario on a fan-spine magnetic topology to explain the observed periodicity.

 In this paper, we report four observations of reflective longitudinal waves seen with HINODE/XRT. One such event is also simultaneously observed by SDO/AIA. We present the observations and the analysis of the events in section~\ref{sec1} and section~\ref{sec2}. The numerical setup and the forward modelling is described in section~\ref{setup} whereas the analysis of the model output data is described in section~\ref{sec4}. Finally we summarize and conclude in section~\ref{sec5}.


\section{Observation and Data reduction}\label{sec1}

The datasets used in this study are obtained from different active regions observed by the X-Ray Telescope (XRT) \citep{2007SoPh..243...63G} on board HINODE \citep{2007SoPh..243....3K} and the Atmospheric Imaging Assembly (AIA) \citep{2012SoPh..275...17L}, onboard Solar Dynamic Observatory (SDO) \citep{2012SoPh..275....3P}.

 The XRT data is calibrated using xrt\_prep.pro (available in SolarSoft package \citet{1998SoPh..182..497F}) which performs the correction for near-saturated pixels, removal of spikes, correction for contamination spots and removal of the CCD bias and the dark current. The final pixel scale and the cadence for each XRT observations are given in Table~\ref{obs_details}.

 The AIA level 1.0 data have been reduced to level 1.5 using the $aia\_prep.pro$ which makes the necessary instrumental corrections. The final pixel scale, in both  X and Y directions, is $\approx$0.6$^{\prime\prime}$. The cadence is 12 seconds.

\section{Data analysis and results}\label{sec2}
\subsection{XRT data analysis}
\subsection{Observation on 10$^{th}$ December 2015}

\begin{figure*}[!htbp]
\begin{center}
 \includegraphics[ angle=90,width=0.80\textwidth]{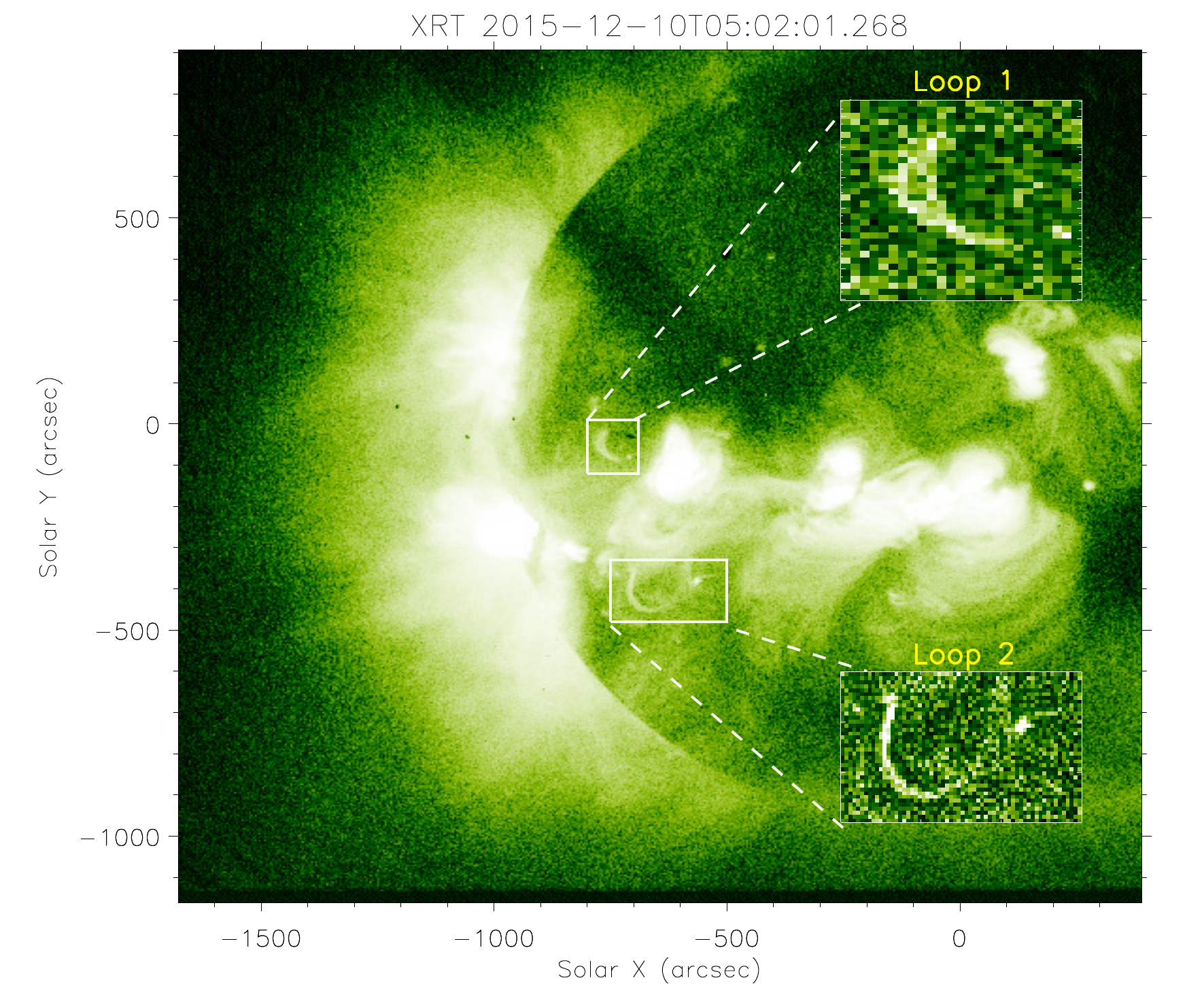}
\caption{The context image showing the full field of view of the observation on 10$^{th}$ December, 2015. Two white boxes highlight the loops of our interest. The zoomed in view of these individual boxes are also plotted on top of the image.}
\label{full_context} 
\end{center}
\end{figure*}

  We used HINODE/XRT data taken in Be-thin filter on 10$^{th}$ December 2015. Figure~\ref{full_context} shows the context image of the observation along with the two loops, (loop 1 and loop 2) from an active region AR 12465, which show reflecting wave propagation.

\begin{figure*}[!htbp]
\begin{center}
 \includegraphics[ angle=90,width=0.70\textwidth]{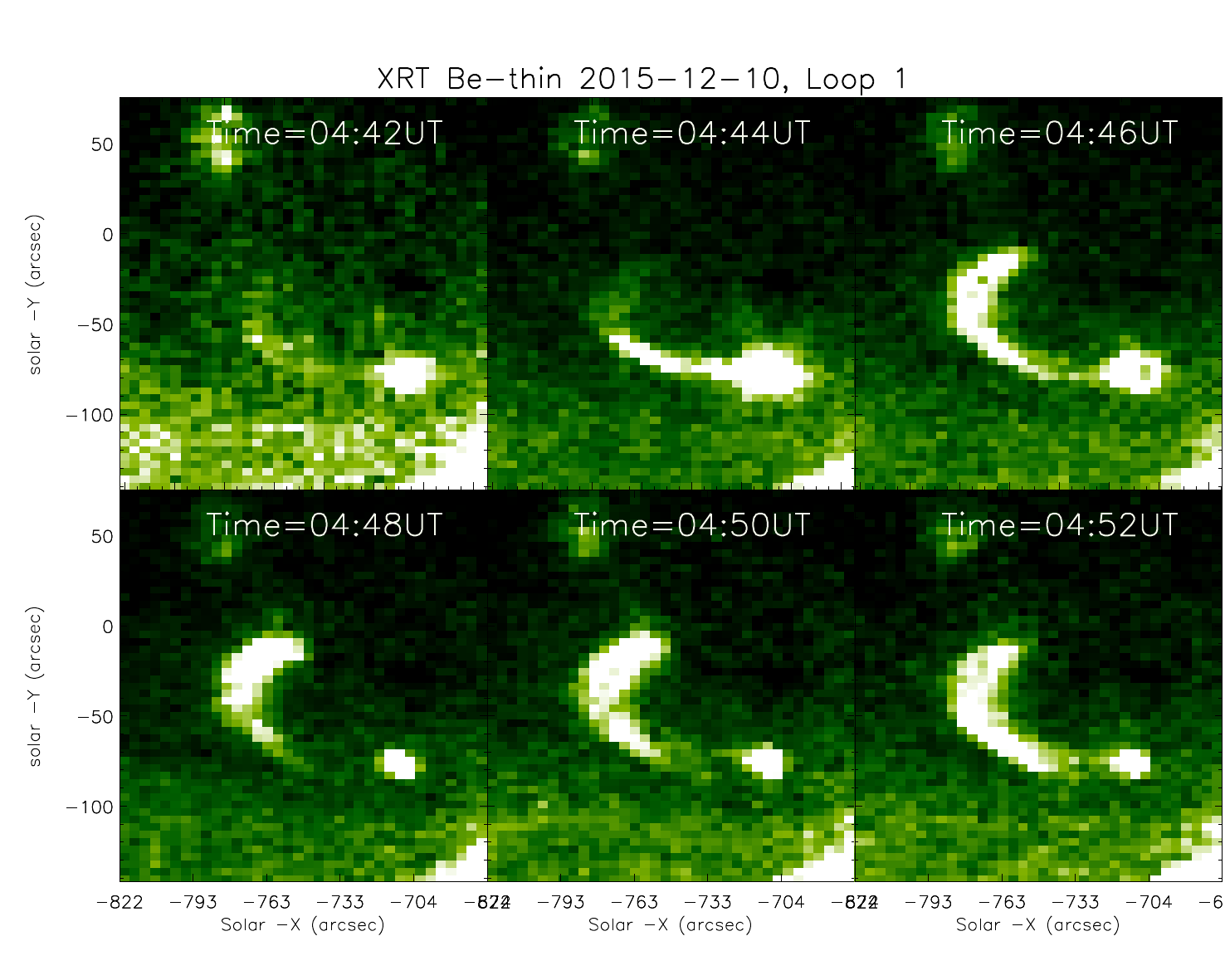}
\caption{ Sequence of base difference images for loop 1. The bright intensity perturbation starts from one footpoint at 04:42 UT and then reflects back from the other footpoint at 04:52 UT. An animated version (movie 1) is available online.}
\label{xrt_new_both} 
\end{center}
\end{figure*}

\begin{figure*}[!htbp]
\begin{center}
 \includegraphics[ angle=90,width=0.93\textwidth]{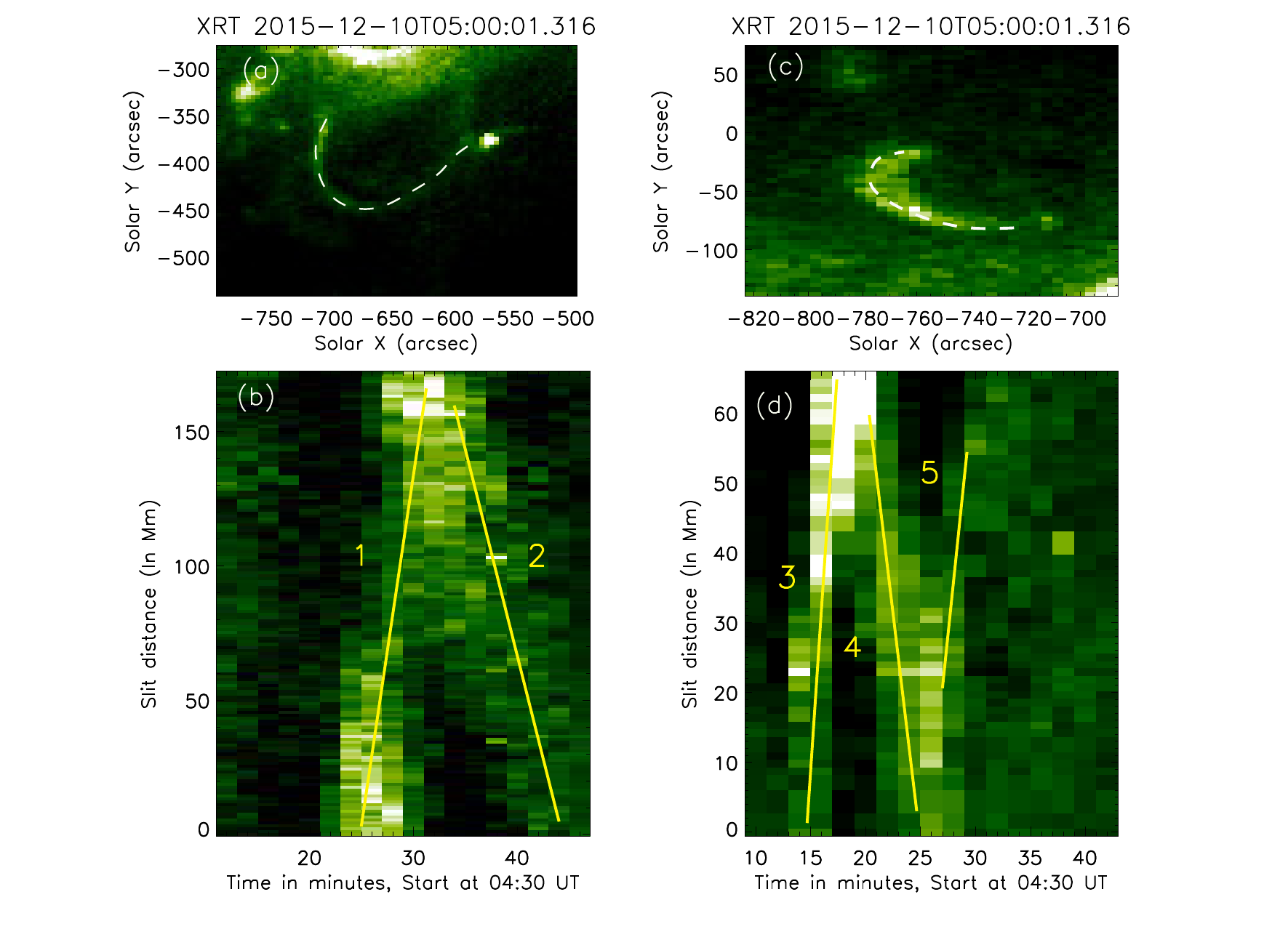}
\caption{(a)-(b) Snapshot of the loop 2 and the obtained histogram equalized time-distance map. The dashed lines show the slits used to generate these maps. (c)-(d) Same as previous but for loop 1. The yellow solid lines (1-5) in both the time-distance maps, highlight the slope of ridges.}
\label{new_event_xt} 
\end{center}
\end{figure*}

 From the movie (movie 1, available online) we see that the onset of the propagating intensity disturbance is caused by a flare which occurred at one of the footpoints. The disturbance then reflects back and forth a couple of times before fading out. To see the propagation of this perturbation through the loop, we stacked different snapshots of the loop 1 in Figure~\ref{xrt_new_both} where one full reflection is seen clearly. We scaled each image to enhance the propagating intensity for better visualization. Speeds of these perturbations are estimated by using the time-distance maps. We put artificial slits tracing the loops to create the time-distance maps. Figure~\ref{new_event_xt} shows the artificial slit positions and the time-distance maps, created using base difference image sequence, for both loop 1 and loop 2. 

For the loop 2 (panel (a) in Figure \ref{new_event_xt}) we see a single reflection within the duration of the observation. The period obtained from the time-distance map (panel (b) in Figure \ref{new_event_xt}) is $~\approx$ 20 minutes. The slopes of the bright ridges are an estimate of the speeds and we obtained a speed of 433 km s$^{-1}$ and 257 km s$^{-1}$ for the ridge 1-2 respectively. For the second loop (loop 2, panel (c) in Figure \ref{new_event_xt}) we see two clear reflections before the signal faded off. The estimated speeds for the ridges 3-5 are 391 km s$^{-1}$, 219 km s$^{-1}$, 251 km s$^{-1}$ respectively. Loop lengths obtained by tracing the loops are 172 $\mathrm{Mm}$ and 61 $\mathrm{Mm}$ for loop 1 and loop 2 respectively. These loop lengths are projected lengths and thus can be an underestimation of the actual lengths.

\begin{figure*}[!htbp]
\begin{center}
 \includegraphics[ angle=90,width=0.85\textwidth]{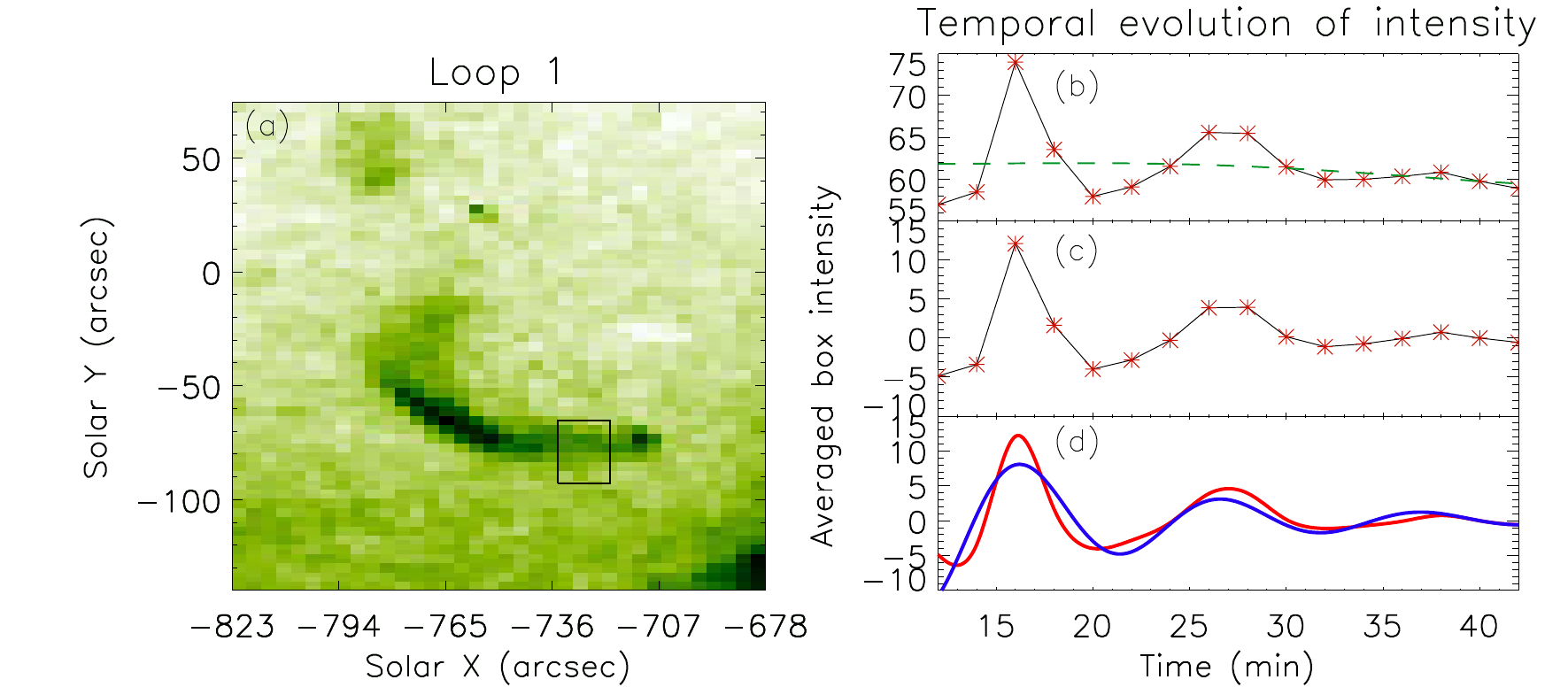}
\caption{(a) Snapshot of the region showing the loop structure (in inverted color). The black box outlines the region selected to extract the intensity. (b) Temporal evolution of the averaged intensity over the box. The small trend, highlighted with a dashed green line, has been subtracted from the original curve to produce the detrended light curve as shown in panel (c). (d) Interpolated light curve is shown with the red line and the fitted decaying sinusoidal is shown in the blue solid line. Start time of these profiles are 04:30 UT. An animated version (movie 2) is available online.}
\label{xrt_damping} 
\end{center}
\end{figure*}

 As we see from the time-distance maps and also from the movie, the intensity amplitudes get damped as they propagate along the loop. To get a quantitative measure of the damping we chose a box as shown in panel (a) in Figure~\ref{xrt_damping} to see the evolution of the averaged box intensity with time. The box is chosen close to the footpoint to avoid the line of sight integration effect because of the loop orientation and also to get a good signal to noise ratio near the footpoint. The averaged box intensity is plotted in panel (b) where each peak corresponds to one reflection.

We see that with each reflection the amplitude is decreasing. A small trend, as obtained
by smoothing the original curve with an increasing window size between 3 to 6 points ( i.e
time frames), is subtracted from the original curve (panel (b)) to produce the detrended curve shown in panel (c) in Figure~\ref{xrt_damping}. To measure the period and the decay time, we first do a 20 fold spline interpolation of the original detrended curve, shown as red solid line in panel (d), to produce a smooth curve. Then we fit the interpolated profile with a damped sinusoidal function of the form

\begin{equation}
\centering
  I(t)=\mathrm{A} \sin(\dfrac{2\pi t}{\mathrm{P}}+\phi) \exp(\dfrac{-t}{\tau})
\label{damp_equ}
\end{equation}
where A,$\mathrm{P}$, $\tau$ and $\phi$ are the amplitude, period, decay time and the phase respectively. The best fitted curve is shown in blue solid line in panel (d) of Fig.\ref{xrt_damping}. The estimated period is 10.1 minutes whereas the damping time is 10.6 minutes.

\subsubsection{Observation on 22$^{nd}$ January 2013}

\begin{figure*}[!htbp]
\centering
 \includegraphics[ angle=90,width=1.\textwidth]{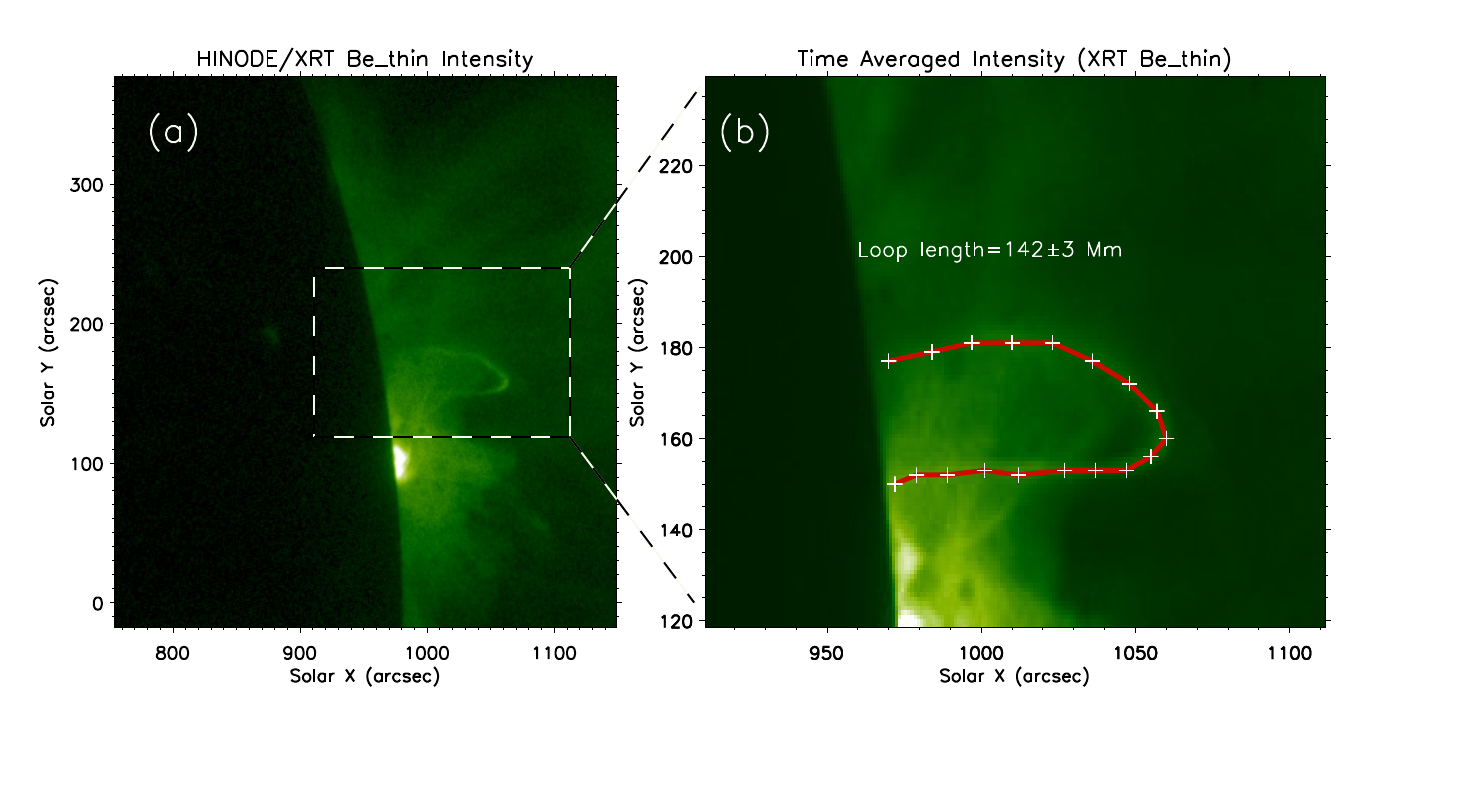}
\caption{Panel (a) shows the full field of view of the 22$^{nd}$ January XRT data. The black-white rectangular box shows the region of interest (ROI) selected for the analysis. A movie of the ROI is available online (Movie~3). Panel (b) shows the time-averaged image of the ROI. The loop length calculated by tracing the loop (white `+' signs) is printed on the plot. The red line is used to create the time-distance plot.}
\label{xrt_context} 
\end{figure*}

We used XRT data taken in Be-thin filter on $22^{nd}$ January 2013 from 8:30 UT to 9:29 UT. The full field of view, shown in panel (a) in Figure~\ref{xrt_context}, is $394'' \times 394''$. The loop of our study is located in active region AR NOAA 11654 in the westward solar limb. We further selected a region of interest (ROI) where the loop is seen clearly and all the analysis has been done on this selected region (panel (b) in Figure~\ref{xrt_context}).

\begin{figure*}[!htbp]
\centering
 \includegraphics[ angle=90,trim={1.5cm 0 0 0},clip,width=1.0\textwidth]{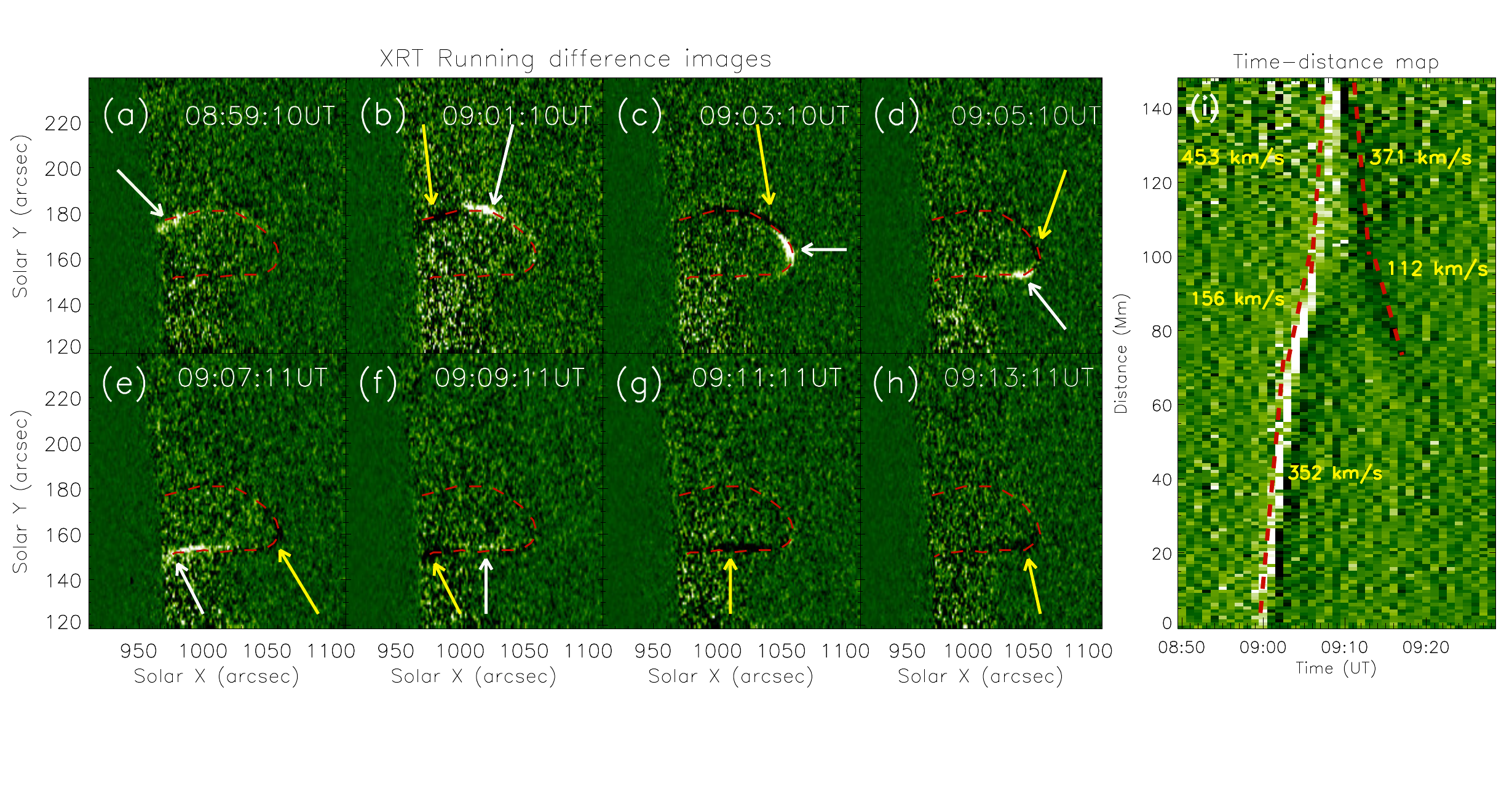}
\caption{Panels (a-h) show the running-difference images created from the XRT Be-thin filter intensity image sequence. The white and yellow arrows show, in each time frame, the current and the previous position of the wavefront respectively. Panel (i) shows the time-distance map from the running-difference image sequence. The white and black inclined ridges show the forward and reflected wave propagation along the loop while the red lines indicate the slope of these ridges. }
\label{xrt_xt} 
\end{figure*}

The wave, appearing from one of the loop footpoints (located on the far-side of the limb) at 08:59 UT, gets reflected from the other footpoint and travels along the loop before fading away. To see the propagation of the wave clearly, running difference images have been created as shown in panels (a-h) in Figure~\ref{xrt_xt}. We see the onset of the propagation at 08:59 UT and we mark it with a white arrow (panel (a) in fig~\ref{xrt_xt}). We also mark the position of the intensity enhancement in the previous frame, seen as a black region following the white, with the yellow arrows. At 9:07UT the wavefront reaches the other footpoint and gets reflected back from there. We have estimated the projected loop length to be 142$\pm3~$$\mathrm{Mm}$ by tracing the points along the length of the loop (the `+' signs in Fig.~\ref{xrt_context}).
 Using the measured loop length and the time the wavefront takes to travel from one footpoint to another (08:59 UT to 09:07 UT), we have an estimate of the average speed of wave propagation and it is $\sim$ 295~$\mathrm{km}$~$\mathrm{s}$$^{-1}$.

 To measure the propagation speed more accurately we have created the time-distance map from the running difference image sequence, as shown in the panel (i) in Figure~\ref{xrt_xt}. The time-distance map has been created using an artificial slit which traces the loop (white `+' signs in panel (b) in Figure~\ref{xrt_context} represent the slit position) and extends the slit width by two pixels across to increase the signal to noise ratio. From the map we clearly see the reflection of the wave from the other end of the loop and then the wave propagates back along the loop before fading away around midway. The slopes of the ridges represent the propagation speeds. Different slopes of the ridges in the time-distance map reflect the change of the loop orientation with the line of sight as the wave propagates along the loop. This event was also co-observed by the `Ti-poly' filter, but neither the loop nor the intensity disturbance was seen in this channel.



\subsubsection{Observation on 27$^{th}$ January 2013}

\begin{figure*}[!htbp]
\centering
 \includegraphics[ angle=90,width=0.79\textwidth]{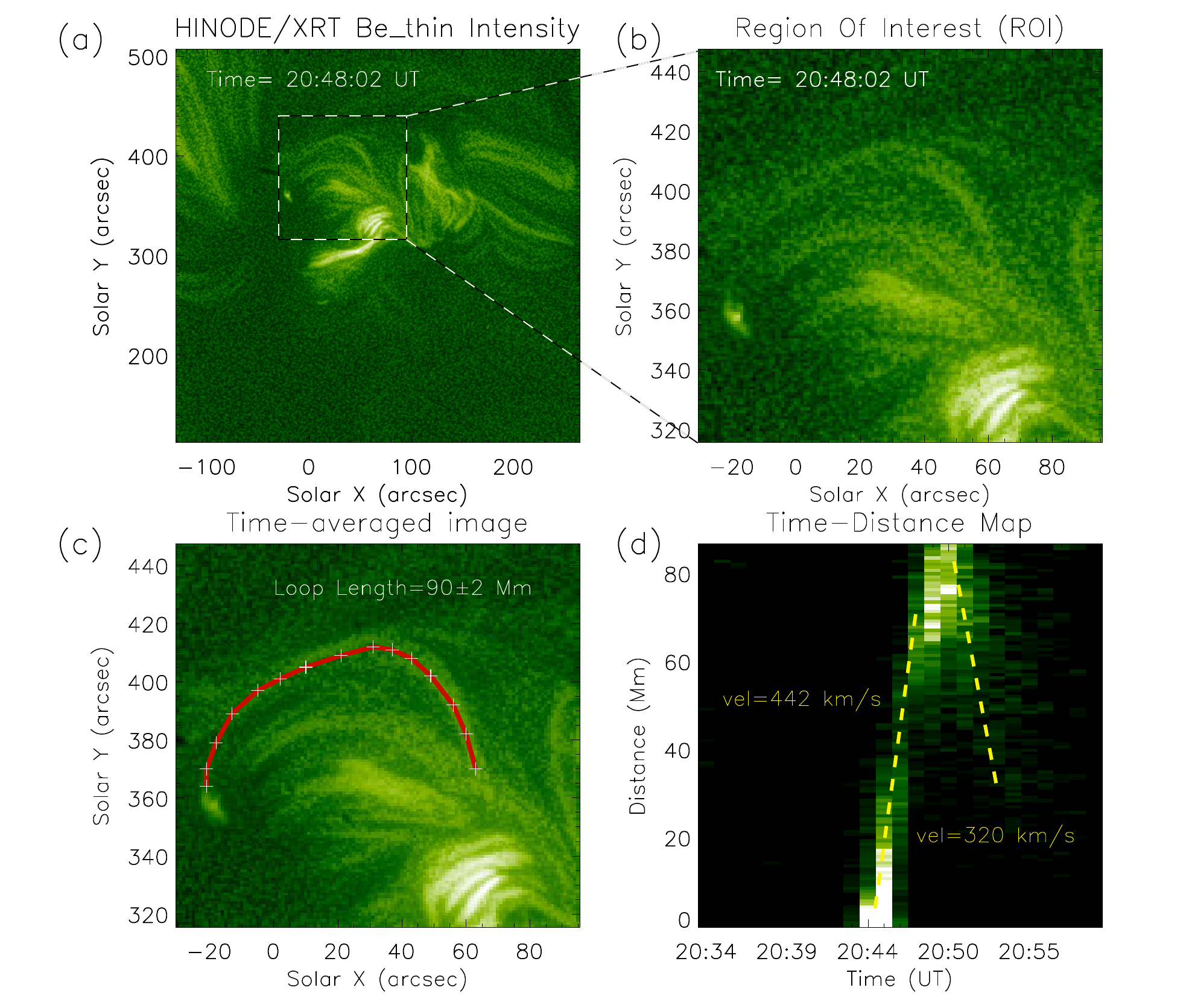}
\caption{ Panel (a) shows the full FOV of the XRT data. The white box indicates the `region of interest'(ROI) that we have selected for the analysis. The zoomed view of the ROI is shown on the panel (b). Panel (c) shows the manually traced loop to calculate the loop length. Processed time-distance map for the traced loop is shown on the panel (d). Estimated speeds are printed on the panel. An animated version (movie 4) is available online.}
\label{context_27}
\end{figure*}

On 27$^{th}$ January 2013, a similar event was observed with the XRT Be-thin filter. A flare like brightening at one of the footpoints is the source of the wave which propagates along the loop. The top panel in Figure \ref{context_27} shows the full FOV of the XRT observation and the region of interest respectively. We have estimated the loop length by tracing points along the loop and the length is equal to 90 Mm (with errors less than 5 Mm). 

From the movie (Movie~4) we see  that the wave is damped rapidly as it propagates through the loop and it faints away as soon as it gets reflected from the other footpoint. This is also seen in the processed time-distance map (bottom last panel in Figure~\ref{context_27}) we created by placing an artificial curved slit following the loop as shown by the `+' signs in Figure~\ref{context_27}. 
The processed time-distance map has been created by scaling individual time in the original map to enhance the bright pixels. In the processed time-distance map we see a reflection signature, though not prominent, from the other footpoint before the signal dropped down considerably. The speed calculated from the inclined ridges (highlighted with yellow dashed line) are 442 $\mathrm{km}$ s$^{-1}$ and 320 $\mathrm{km}$ s$^{-1}$ respectively.

\begin{figure*}[!htbp]
\begin{center}
 \includegraphics[ angle=90,width=0.75\textwidth]{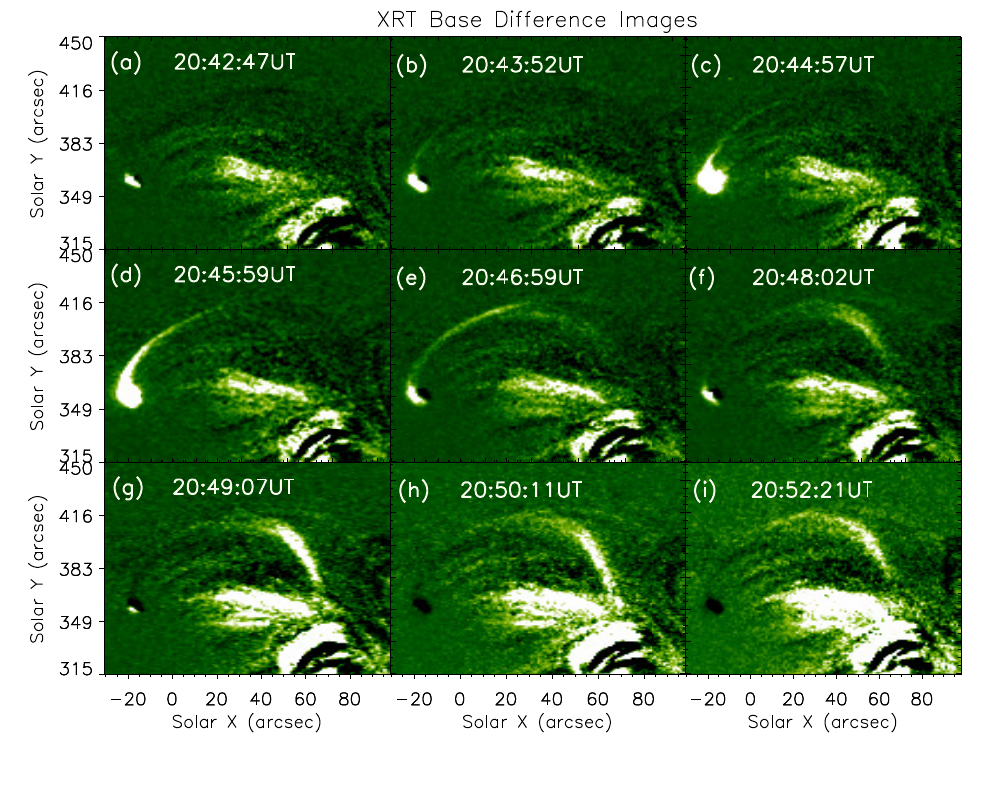}
\caption{Base difference XRT images of the ROI, showing the wave propagation from one loop footpoint to the other. Each image is scaled individually to highlight the wavefront.}
\label{base_27} 
\end{center}
\end{figure*}

The propagation and the reflection of the wave is prominently visible in the base-difference images shown in Figure~\ref{base_27}. Panel (a-h) shows the intensity propagating from the one footpoint to other and in panel (i) we see the reflected part moving in the other direction.

\subsection{AIA data analysis}

 The XRT `Be-thin' filter mostly observes the `hot' plasma (the response function of `Be-thin' filer peaks at 10$\mathrm{MK}$). This leads to the fact that the event observed in this filter is likely to be captured in hotter channels of AIA ( 94~\r{A} and 131~\r{A}). Despite of this, we found only one event (on 22$^{nd}$ January, 2013) where the loop is also simultaneously seen with AIA. This is because the AIA 94~\r{A} and AIA 131~\r{A} channels have a second peak around 1.5~$\mathrm{MK}$ and 0.5~$\mathrm{MK}$ respectively, next to their expected response at 10$~\mathrm{MK}$. These secondary peaks contaminate heavily when the loop is on disc rather than when it is on the limb. Also for the other events, the plasma may be too hot to be seen in any of the AIA channels. It must be mentioned here that though we did not see the loop (except for the event on 22$^{nd}$ January, 2013), we have captured the time evolution of the loop footpoints for all the other events. For the event on 22$^{nd}$ January, 2013 we could not locate the footpoint as it was on the far side of the Sun.

\begin{figure*}[!htbp]
\centering
 \includegraphics[ angle=90,trim={1.5cm 0 0 0},clip,width=0.8\textwidth]{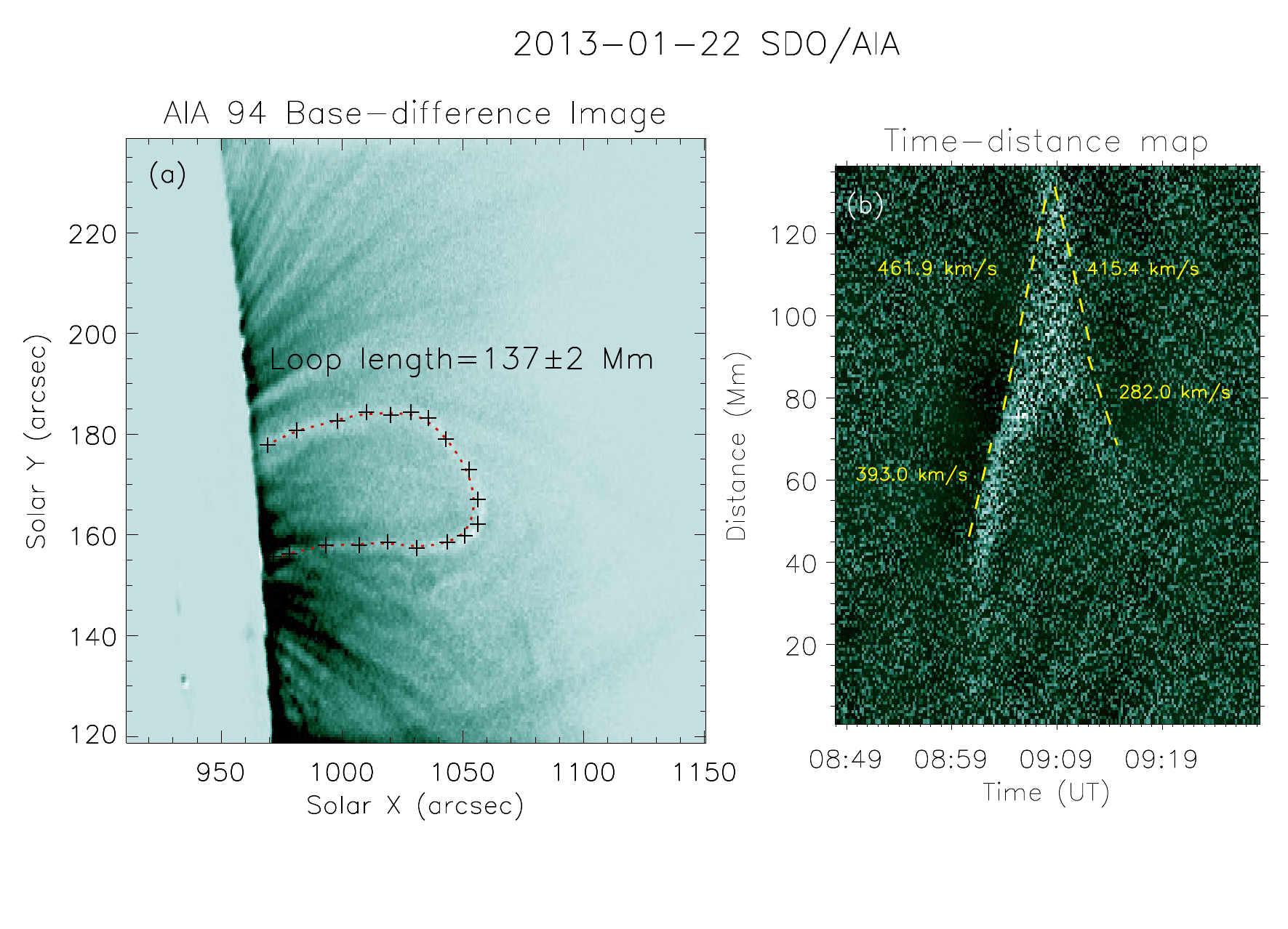}
\caption{(a) AIA 94~\r{A} time averaged image created from the base-difference image sequence. The detected loop is marked by `+' signs. (b) The time-distance plot showing two oppositely inclined white ridges. Yellow lines indicate the slope of these ridges. An animated version (movie 5) is available online.}
\label{aia_xt} 
\end{figure*}
 For the event on 22$^{nd}$ January 2013, we found that the loop is also detected in the AIA 94~\r{A} channel (only) indicating the presence of a high temperature plasma in the observed loop. We have used corresponding  SDO/AIA data in the 94~~\r{A} channel taken from 8:30 UT to 9:30 UT. We used this data to co-align the two instruments and the final XRT coordinates are obtained after correcting for the offsets. The left panel in Figure~\ref{aia_xt} shows the time averaged image of base-difference image sequence of the AIA 94~~\r{A} channel. We also estimate the loop length in a similar way as the XRT measurement and the length is equal to 137$~\pm~$2 Mm. From the movie (movie 5, available online) we clearly see that the wave starting from one footpoint gets reflected from the other footpoint and  faints away after traveling a certain distance along the loop-length. This feature is also seen clearly from the time-distance map created from the AIA base-difference image sequence. Two oppositely inclined ridges are visible in the map indicating a clear reflection signature. We also estimated the propagation speed from the slope of the ridges and they are shown in Figure~\ref{aia_xt}.

\subsection{DEM analysis}

Wave propagation through a loop largely depends upon the physical parameters of the loop density, temperature etc. Since the sound speed in a medium depends on its temperature, we performed automated temperature and emission measure analysis, using a SolarSoft code, as developed by \citet{2013SoPh..283....5A} to estimate average density values and temperature inside the coronal loop.
\begin{figure*}[!htbp]
\centering
\includegraphics[width=0.80\textwidth]{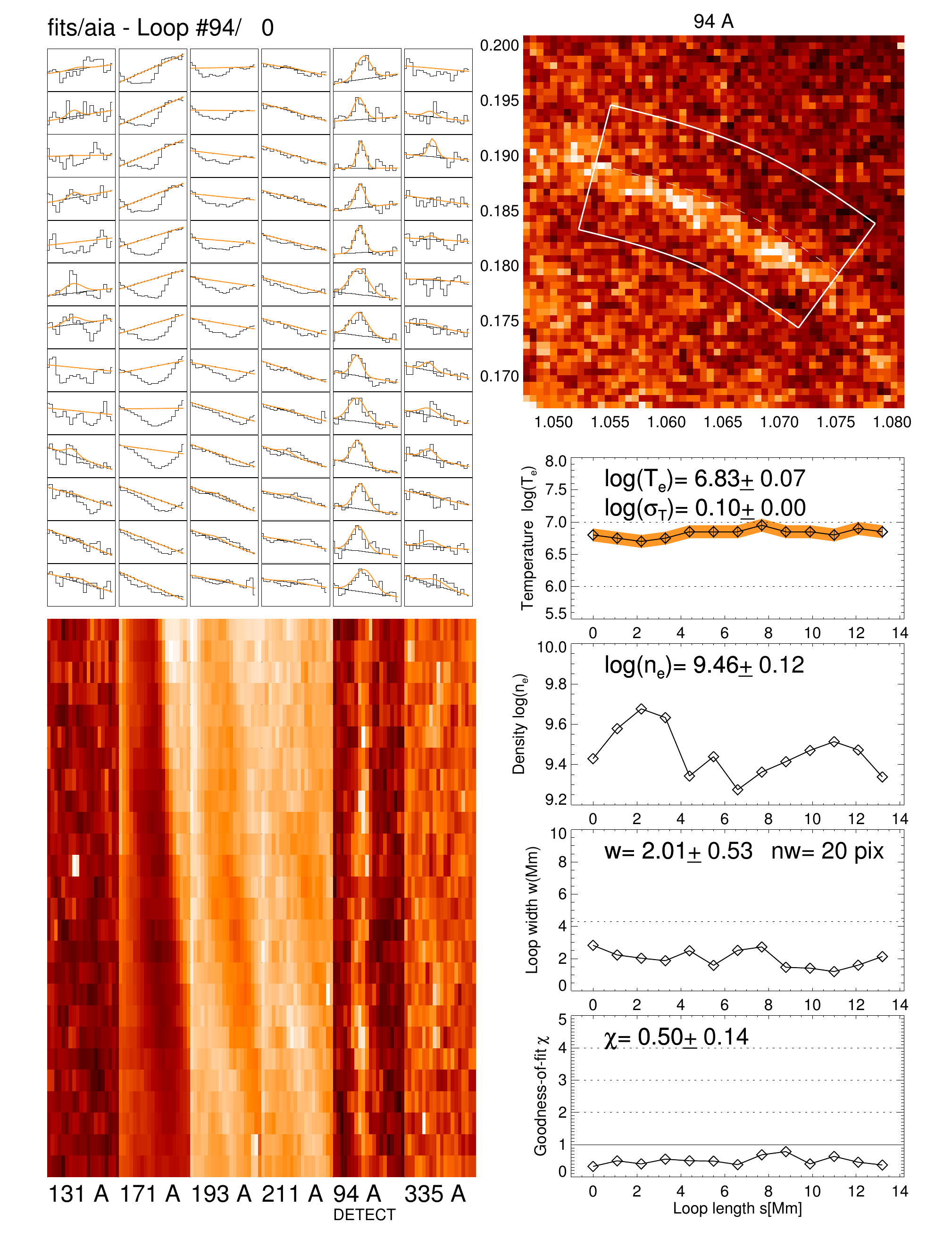}

\caption{Automated DEM analysis performed with the AIA dataset. The loop is only seen in the AIA 94~\r{A}. The top left panels show the Gaussian fits performed on the selected loop structure (top right panel) to estimate the loop width. The obtained loop density (n$_\mathrm{e}$), loop temperature (T$_\mathrm{e}$) and the goodness of fit ($\chi$) is also plotted in different panels.}
\label{aia_dem} 
\end{figure*}

\begin{figure*}[!htbp]
\centering
\includegraphics[trim={10.5cm 0 0 0},clip,width=0.4\textwidth]{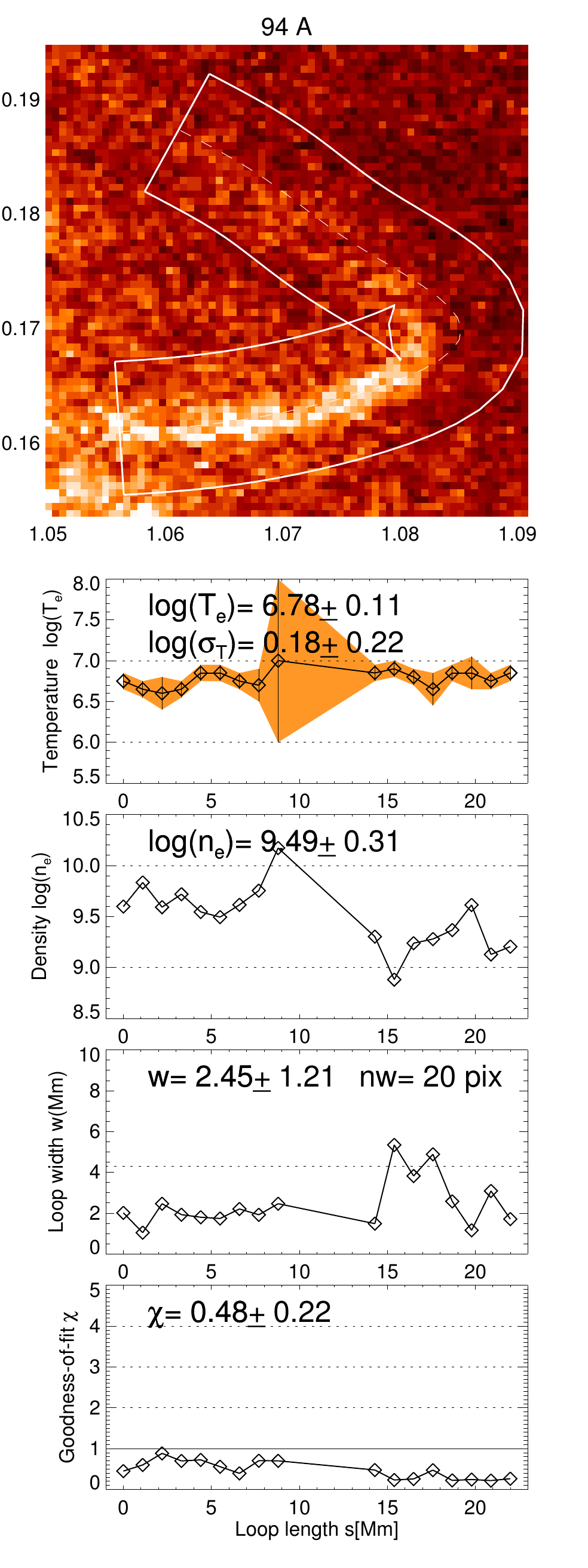}
\includegraphics[trim={10.5cm 0 0 0},clip,width=0.4\textwidth]{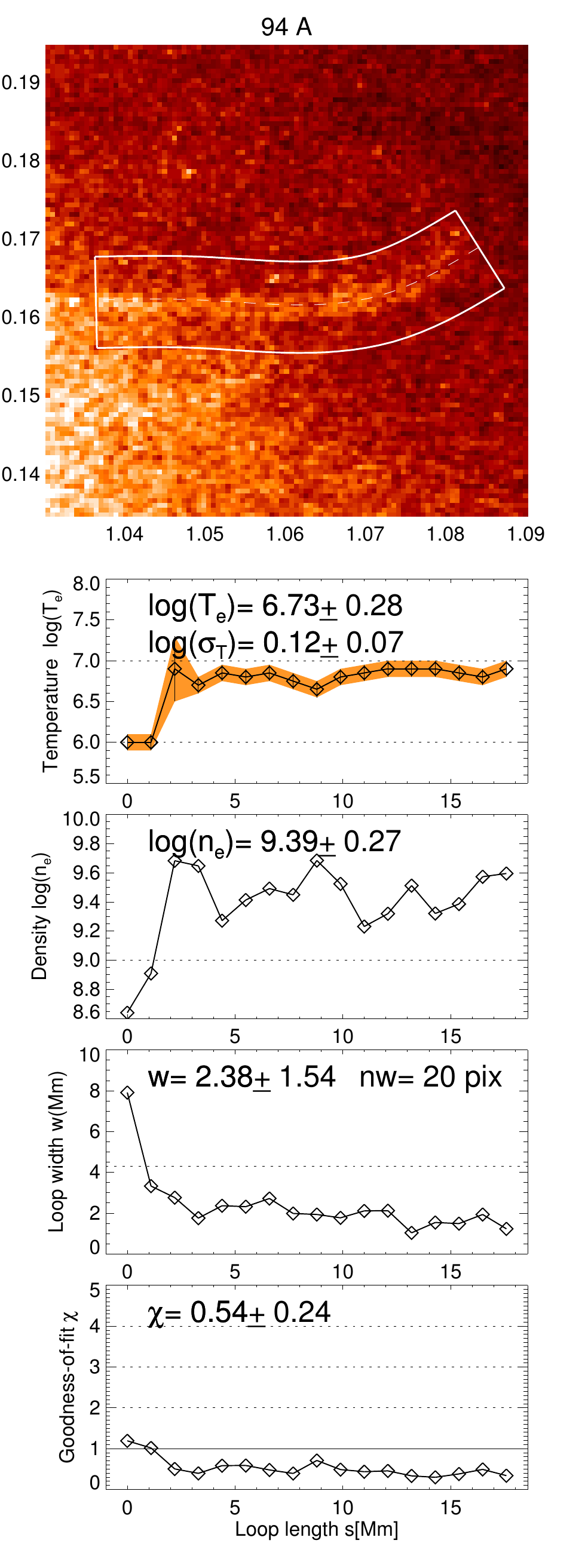}
\caption{Automated DEM measurements same as Figure~\ref{aia_dem} but for two different location at the loop. The error values are derived from the width of the Gaussian fitting (see \citet{2013SoPh..283....5A} for details) and may be an underestimation of the actual uncertainties.}
 \label{dem13}
\end{figure*}

 In order to estimate the DEM, the SolarSoft code co-alignes AIA images in different EUV passbands (94~\r{A}, 131~\r{A}, 171~\r{A}, 193~\r{A}, 211~\r{A}, 335~\r{A}) using solar limb fit. Using the forward fitting of the observed DEM distribution with that of the model DEM distribution, the peak emission measure and the peak temperature at a particular pixel is determined. In our case, the loop is only poorly visible in AIA 94~\r{A} channel. Thus the density and temperature values, obtained from the DEM analysis, are only rough estimations for both the parameters. This order of magnitude estimate is sufficient for our purpose as we insert these values in our numerical model (described in subsequent section) as initial loop parameters. To check for any spatial as well as temporal evolution of the loop parameters, we performed DEM analysis at three different locations at three different times. One such case is shown in detail in Figure~\ref{aia_dem}. The estimated average loop density and temperature are $\approx$10$^9~$cm$^{-3}$ and $\approx$10 MK respectively. Using this temperature value, the sound speed within the loop is c$_s$$\sim$ 152$\sqrt{T(\mathrm{MK})}$$\sim$480~$\mathrm{km}$~$\mathrm{s}^{-1}$. The other two DEM measurements are shown in Figure~\ref{dem13}.

\section{Generation Mechanism}
 
 The generation mechanism of such waves is not fully understood yet. There are evidences of small (-micro) flares, occurring at the loop footpoints, as seen in our events also, to be the generator of the waves. See \citet{2011SSRv..158..397W} for a complete review on this.

\begin{figure*}[!htbp]
\begin{center}
 \includegraphics[ angle=90,width=0.90\textwidth]{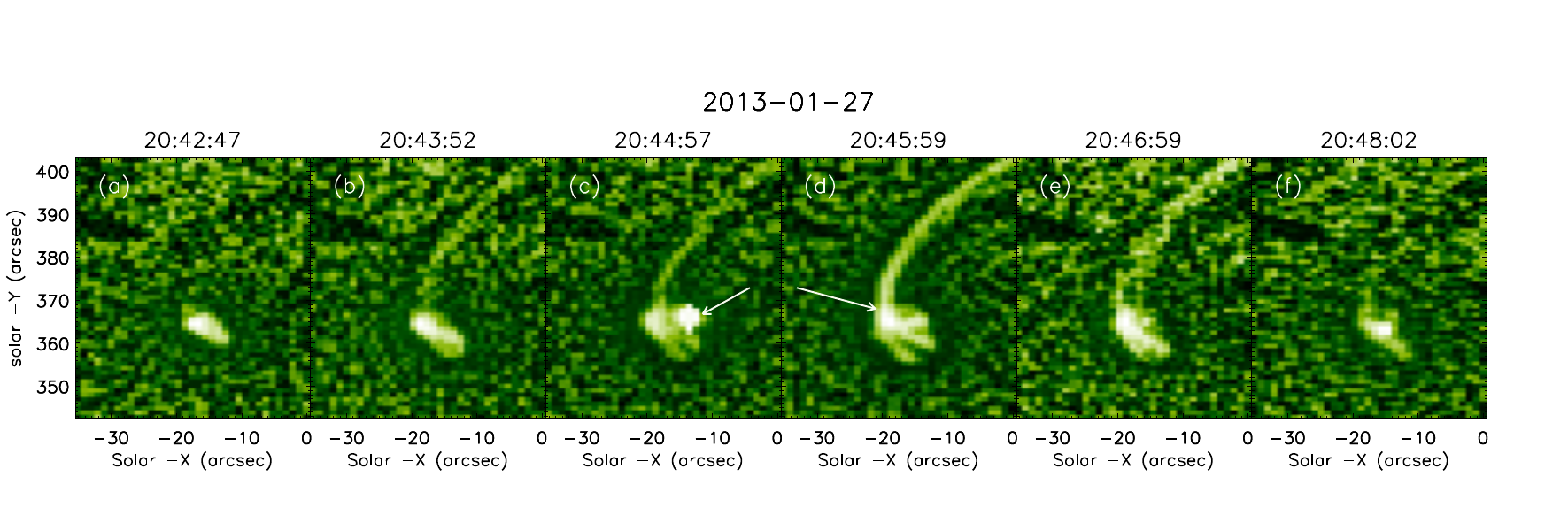}
\caption{ Snapshots showing the loop footpoint and its temporal evolution, for the event on 27$^{th}$ January, as seen by the XRT Be-thin filer. All the images have been scaled to highlight the footpoint. The white arrows in panels (c) and (d) indicate the brightening position in the corresponding frames.}
\label{footp_27_xrt} 
\end{center}
\end{figure*}

 In the event on 22$^{nd}$ January, the flare happened at the far side of the Sun and thus it was hidden from us (STEREO also did not capture the event due to corser cadence of 5 minutes). We did capture the loop footpoint for the event on 27$^{th}$ January from XRT as well as from AIA EUV channels. Figure~\ref{footp_27_xrt} shows the time evolution of the loop footpoint as seen with XRT. We notice a complex mini-loop like structures at the footpoint with many spines as revealed in the image sequence of Figure~\ref{footp_27_xrt}. We also notice that an intense brightening which starts at the right end of the loop footpoint (panel (c)) moves towards the left in the next time frame (panel (d)). This observed occurrence and evolution of the brightening at the loop footpoint resembles very much the `blowout-jet' examples found previously \citep{2013ApJ...769..134M}.

\begin{figure*}[!htbp]
\begin{center}
 \includegraphics[ angle=90,width=0.92\textwidth]{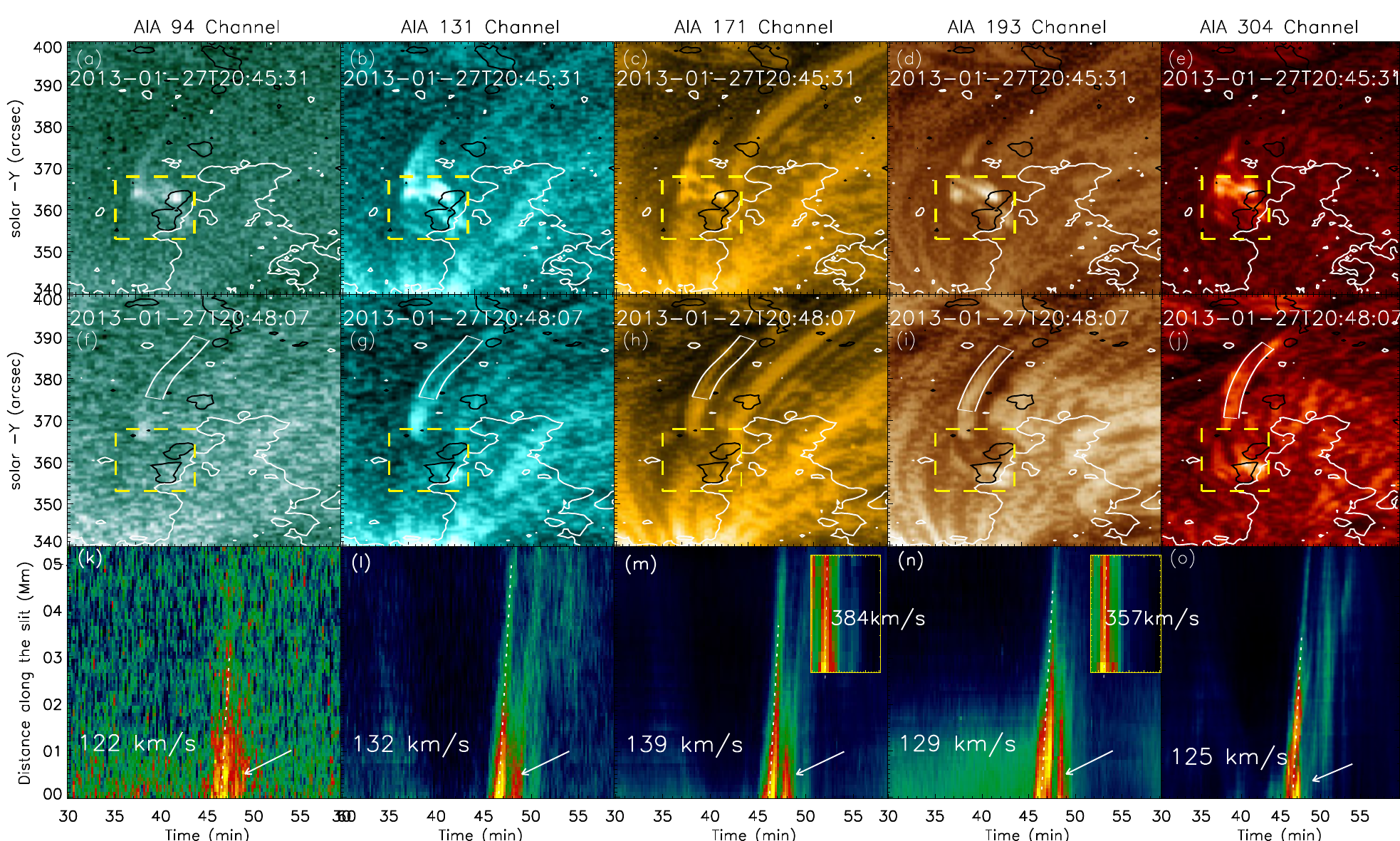}
\caption{ Panels (a-e) show the snapshots of the footpoint at the peak flare time in five AIA EUV filters (171~\r{A}, 131~\r{A}, 94~\r{A}, 193~\r{A}, 304~\r{A}). The black and white contours represent the HMI LOS negative and positive fields ($\pm$20 G). The yellow rectangular box covering the footpoint is used create the light curved shown in Figure~\ref{lc}. Panels (f-j) show the same region at a later time. White rectangles highlight the position of the artificial slit used to generate time-distance maps (histogram equalized for better visualization) shown in panel (k-o). Speeds measured by calculating the slope of the white dotted line, are printed in each panel. A second peak in the time-distance map is highlighted by a white arrow in each panel ((k)-(o)). Speed measurements for this second peak is also shown in insets (panel (m)-(n)). Start times of the time-distance maps are 08:00 UT.}
\label{footp_27_aia} 
\end{center}
\end{figure*}

Though the loop is not visible in any of the AIA channels but the loop footpoint is clearly visible in all the EUV channels. Figure~\ref{footp_27_aia} shows the time evolution of the footpoint in five EUV channels of AIA (171~\r{A}, 131~\r{A}, 94~\r{A}, 193~\r{A}, 304~\r{A}). Panels (a-e) show the snapshots of the footpoints at the time when the flare peaks. We see a complex loop arcade with a spine forming at the top of the structure. From the movie (movie 6) we also notice that the AIA cooler channels (171\r{A} and 304\r{A}) along with a hotter channel (193\r{A}) capture a small `filament like' dark feature rising with the evolution of the flare {\citep{2015Natur.523..437S}}. Though we must emphasize the fact that the initial location of the filament before the flare is not clearly visible. To see the photospheric magnetic filed configuration associated with this structure, we overlay the Helioseismic and Magnetic Imager (HMI) line of sight (LOS) magnetic filed contours ($\pm$ 20 G) on top of the AIA images. We do not see a clear bipolar structure around the footpoint region in this case suggesting the fact that the reconnection might have happened higher in the atmosphere. Panels (f-j), in Figure~\ref{footp_27_aia}, show the snapshots of the same region when the ejecta is seen to propagate through the spine. Though the ejected material is poorly visible in the AIA 94~\r{A} channel but it shows up in rest of the channels. A yellow rectangular box, which marks the footpoint region in panels (a-e), is also overplotted in panels (f-j) for better comparison.

          The footpoint structure almost disappeared at this moment (panels f-j) leaving the spine through which the ejected plasma moves. To measure the propagation speed of the plasma, in the plane of sky, we create the time-distance maps for all the channels by placing an artificial slit as indicated by a white curved rectangle in panels (f-j). The obtained time-distance maps for individual channels are plotted in the bottom panel of Figure~\ref{footp_27_aia} (panels (k-o)). Here we see two slanted ridges in all the channels except the 304~\r{A} channel where the second ridge is not visible. The speeds, calculated from the slopes of the first ridge in all the cases, ranges from 122-139 $\mathrm{km}$~s$^{-1}$.

\begin{figure*}[!htbp]
\begin{center}
 \includegraphics[ angle=90,width=0.70\textwidth]{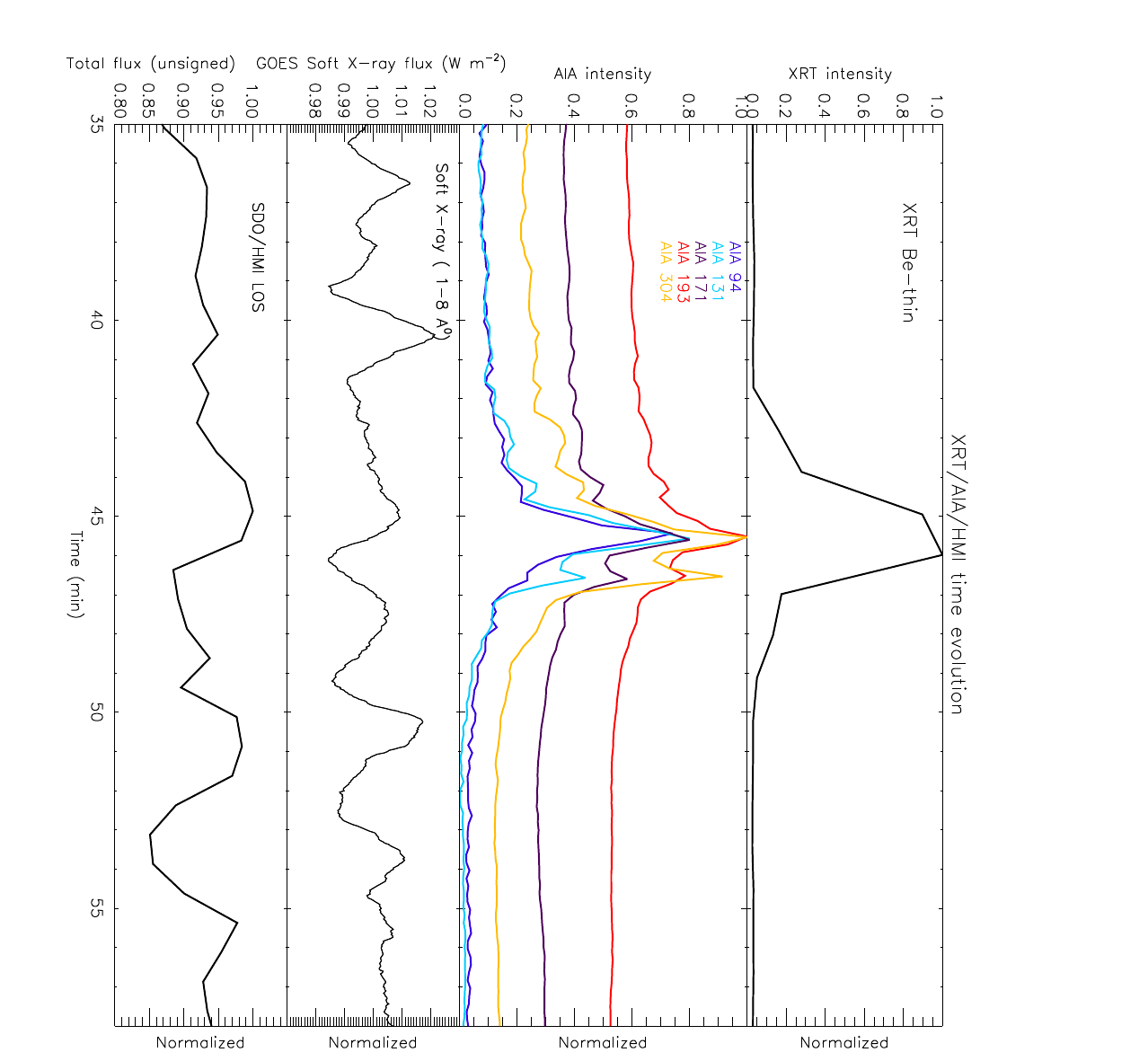}
\caption{ Top panel shows the time evolution of the box averaged intensity for the XRT data. AIA intensities for the five AIA channels, are plotted in the middle panel. Bottom two panels show the light curve for GOES soft X-ray flux (1-8\r{A}) and the HMI unsigned total flux (averaged over the box). Start times of the profile are 08:00 UT.}
\label{lc} 
\end{center}
\end{figure*}
 The occurrence of this ejecta coincides with flare peak time (at 20:45 UT) as seen from XRT . The second ejecta, which occurred $\approx$ 2.5 minutes later, propagates with a much faster velocity than the first ejecta. The measured speeds in this case are 384 $\mathrm{km}$~s$^{-1}$ and 354 $\mathrm{km}$~s$^{-1}$ in 171~\r{A} and 193~\r{A} channels respectively.

 To understand the evolution of the observed footpoint intensity and its association with the photospheric magnetic field, we plot in Figure~\ref{lc}, the temporal evolution of the intensities from AIA and XRT and the unsigned HMI magnetic flux averaged over the yellow box shown in panel (a) of Figure~\ref{footp_27_aia}. GOES soft X-ray flux (1-8 ~\r{A}) is also in the plot. We see a clear association of the XRT intensity peak with intensity increments in all the AIA EUV channels. Also this increment is accompanied by a soft X-ray flux enhancement. A closer investigation on the AIA light curves shows another peak, approximately 1 minutes later, with smaller magnitude than the first one. It is worth to mention here that the second peak is very weak in 94~\r{A} channel and there is no such peak seen in XRT intensity prfile. The absence of the second peak in the XRT profile can be explained by considering the coarser time and spatial resolution of the XRT compared to the AIA and also the filter response of the XRT `Be-thin' filter. In the last panel of Figure~\ref{lc} we plot the total (unsigned) flux from the boxed region and we notice that there are two dips on the profile around the flare time (though these dips are not significantly strong and compareable to the fluctions seen in the time series). Note the second ridge, found in the time-distance maps in Figure~\ref{footp_27_aia}, originates due to the second peak found in the intensity profiles of AIA. We propose a scenario where the chromospheric plasma gets heated rapidly due to the flare and produces the high speed ejecta which is seen to propagate in all the coronal channels. Here we want to highlight the fact that there is a delay of $\approx$1.5 minutes between the second intensity enhancement and occurrence of the high speed plasma. Such a scenario of rapid heating of the chromospheric material also explains the absence of this second ejecta in chromospheric 304~\r{A} channel.

\begin{figure*}[!htbp]
\begin{center}
 \includegraphics[ angle=90,width=0.98\textwidth]{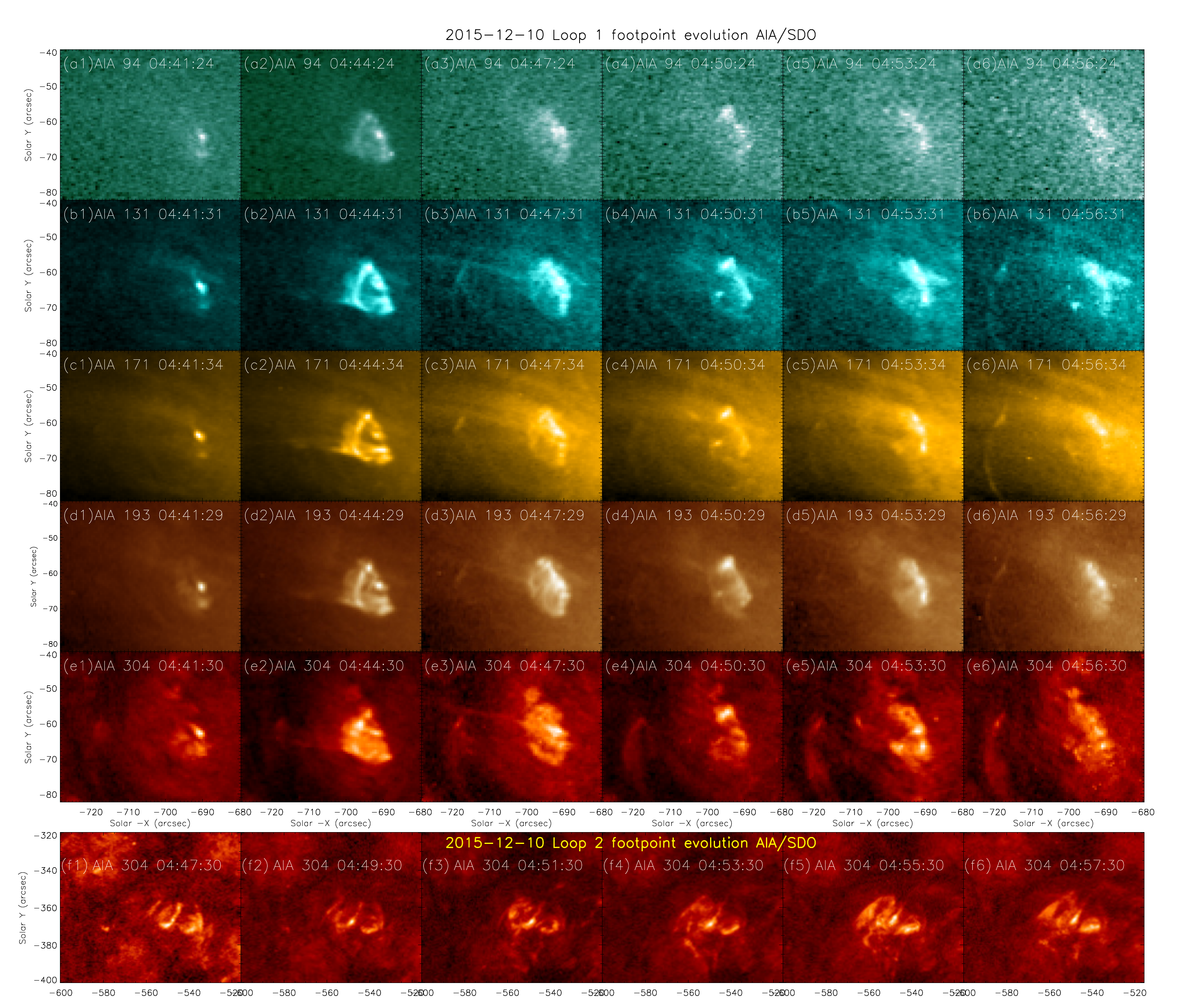}
\caption{ Panels (a1-e6) display the time evolution of the loop footpoint for the event on 10$^{th}$ December, 2015 in five EUV channels during the flare time. Panels (f1-f6) show the same but for footpoint of loop2.}
\label{aia_new_foot} 
\end{center}
\end{figure*}

\begin{figure*}[!htbp]
\begin{center}
 \includegraphics[ angle=90,trim={5cm 1cm 0 0},clip,width=0.92\textwidth]{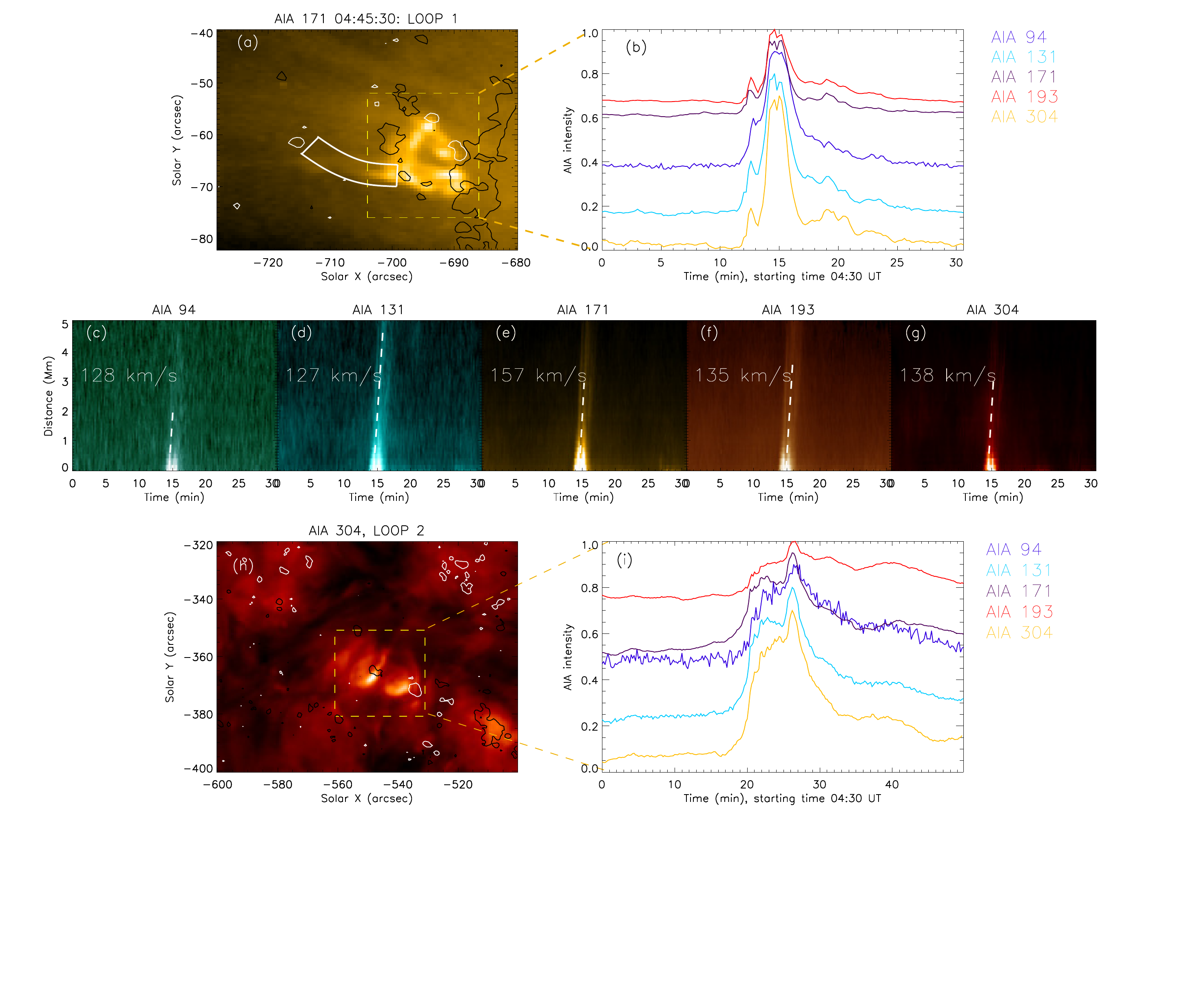}
\caption{Panel (a) shows the snapshot of the loop1 footpoint in AIA 171~\r{A} channel. The yellow box highlights the region used to calculate the footpoint intensity evolution in five AIA EUV channels as shown in panel (b). The white rectangular box in panel (a) marks the artificial slit used in creating time distance maps (panel (c-g)). Panel 
(h) shows the footpoint of the loop2 and the corresponding intensity evolution (within the yellow box) is presented in panel (i). The black and white contours in panel (a) and (h) represent the negative and positive polarities of the LOS magnetic fields ($\pm$30G) as obtained from HMI.}
\label{fig:15} 
\end{center}
\end{figure*}

 Now we analyze the event occurred on 10$^{th}$ December, 2015. XRT has observed this event with 4$\times$4 spatial binning, resulting in a spatial scale of 4.1$ ''$ in both x and y directions. Thus we could not resolve the footpoint from the XRT data (the cadence was also coarser in this case, 121 seconds). Using the AIA data, we do not see the loop structure in this case too but the footpoints (for both loop1 and loop2) are seen in all the AIA EUV channels. Panels (a1-e6), in Figure~\ref{aia_new_foot}, show the temporal evolution of the loop1 footpoint in five EUV channels of AIA. Similar to the previous event, here also we see a jet spine at one side of the footpoint. Snapshots of the footpoint, for the loop2, are shown in panels (f1-f6). In this case, we do not see a clear jet spine structure like the previous events though we see multiple `mini-loops' at the footpoint similar to the other events. Figure~\ref{fig:15} shows the temporal evolution of the footpoint intensities for both the loops. Panel (a) of Figure~\ref{fig:15} highlights the yellow rectangular box chosen for the intensity evolution study of the footpoint and also highlights the artificial slit (white rectangular box) used to create the time-distance maps (shown in panels (c-g) in Figure~\ref{fig:15}). From the time evolution of the intensities in different AIA channels, shown in panel (b), we notice that a major peak occurs at 04:45 UT which matches with the onset of the wave propagation as seen by XRT ( panel (d) in Figure~\ref{new_event_xt}). Now, from the time-distance maps (panels (c-g)) we see that an inclined ridge appears, in all the AIA channel, co-temporally with the intensity peak. The speeds of propagation, as measured from the slope of the ridges, range between 127-157 $\mathrm{km}$ sec$^{-1}$. We have also overplotted the HMI LOS magnetic field contours ($\pm$30 G) in panel (a). We see the presence of both positive and negative polarities within the yellow box but the dioplar structure is not clearly visible. For the footpoint of loop2 (panel (h)), we do not see any clear jet or spine structure but the footpoint evolution of the intensities, in all the channels, shows a peak at 04:27 UT which again matches with the onset of the wave propagation seen from XRT ( panel (b) in Figure~\ref{new_event_xt}). We could not use the GOES data in this case due to the presence of multiple active regions and other co-temporal events occurring on the disc (close to the event location).

 In summary, we see that a micro-flare at one of the footpoints of the loops, acts as a trigger for the slow waves. The micro-flare originally ejects a plasma which, as soon as it becomes detached from the source, evolves as a wave packet and exhibits slow wave properties. We test our idea of such flare generating the slow waves in our numerical model (described in next section) by injecting heat equivalent to such flare energy at one of the loop footpoint.

\section{Numerical Experiment}\label{setup}

From the observations we obtained an estimate about the speed of the wave propagating through the loop. The the density and temperature values of the loop plasma are also calculated using DEM analysis using the AIA data. Now we use a numerical simulation with the obtained loop length, density and temperature as the input parameters, to model the observations. Our simulation uses a 2.5D thermodynamic magnetohydrodynamic model as in \citet{2015arXiv150904536F} which includes gravity, anisotropic thermal conduction and radiative cooling. The box domain in the simulation is taken as -60 Mm $\leq  x \leq 60$ Mm and 0 $\leq  y \leq 80$ Mm in order to obtain a comparable loop length of $\approx$140 Mm as estimated from our observations.

We initialize with a linear force-free magnetic field given by
\begin{displaymath}
 B_{x}=-B_{0} \cos \left( \frac{\pi x}{L_{0}} \right) \sin\theta_0 \exp\left(
 -\frac{\pi y \sin\theta_0}{L_{0}} \right)\,,
\end{displaymath}
\begin{displaymath}
 B_{y}=B_{0} \sin \left( \frac{\pi x}{L_{0}} \right) \exp\left(
 -\frac{\pi y \sin\theta_0}{L_{0}} \right)\,,
\end{displaymath}
\begin{equation}
 B_{z}=-B_{0} \cos \left( \frac{\pi x}{L_{0}} \right) \cos\theta_0 \exp\left(
 -\frac{\pi y \sin\theta_0}{L_{0}} \right)\,,
\label{bfield}
\end{equation}
with the angle $\theta_0=30^\circ$ between the arcade and the neutral line ($x = 0, y=0$) and the horizontal size of our domain setting $L_{0}=120$ Mm (-60 Mm $\leq  x \leq 60$ Mm) and we take $B_{0}=$ 50 $\mathrm{G}$. We set the temperature below a height of 2.7 Mm as a uniform 10,000 K for the initial thermal structure. The distribution of initial density is calculated based the assumption that a hydrostatic equilibrium with a number density of 1.2 $\times$ 10$^{15}$ cm$^{-3}$ lies at the bottom of the simulation box. We assume the initial setup with a background heating rate which decays exponentially with height for approaching a self-consistent thermally structured corona, $H_{0}=c_{0} \exp\left(-\frac{y }{\lambda_{0}} \right)$ where $c_{0}=10^{-4}$ erg cm$^{-3}$ s$^{-1}$ and $\lambda_{0}=80$ Mm. This heating is used to balance the radiative losses and anisotropic heat conduction related losses of the corona in its equilibrium state. With this initial setup, we integrate the governing MHD equations until the above configuration reaches a quasi-equilibrium state, when we reset time to zero.

 We use the MPI-parallelized Adaptive Mesh Refinement Versatile Advection Code  $\mathrm{MPI-AMRVAC}$ \citep{2012JCoPh.231..718K,2014ApJS..214....4P} to run the simulation. An effective resolution of $1536 \times 1024$ or an equivalent spatial resolution of 79 km in both directions is obtained through four AMR levels.
The energy release from the flare is mimicked by a finite duration heat pulse $H_{1}$ located at the right footpoint between $x=39, 40$ Mm.
\begin{equation}
\begin{array}{lrrr}
 H_{1}=c_{1} \exp(-(y-y_{c})^{2}/\lambda^2)f(t) & {\mathrm{ if }} & {A(x_{1},0)<A(x,y)<A(x_{2},0)}
 \end{array} 
\end{equation}
\begin{equation}
A(x,y)=\frac{B_{0}L_{0}}{\pi}\cos\left(\frac{\pi x}{L_{0}}\right)\exp\left(-\frac{\pi y \sin\theta_{0} }{L_{0}} \right) \,,
\end{equation}
\begin{equation}
f(t)=\left\{
\begin{array}{lrr}
t/30 & 0\leq t <30 &{\mathrm{s}}\\
1 & 30\leq t <150 &{\mathrm{s}}\\
(180-t)/30 & 150\leq t <180 &{\mathrm{s}}
\end{array} \right.
\end{equation}
where $\lambda^{2}=10$ Mm$^{2}$, $x_{1}=40$ Mm, $y_{c}=3$ Mm  and $x_{2}=39$ Mm.
 The pulse is switched on only for a time $t=0$ to $t=180$ seconds. we set the $c_{1}$ to 12 erg cm$^{-3}$ s$^{-1}$ in our simulation input. We introduced the anisotropic thermal conduction along the magnetic field lines with the Spitzer conductivity $\kappa_{||}$ defined as 10$^{-6}T^{5/2}$ erg cm$^{-1}$ s $^{-1}$ K $^{-3.5}$ . We use radiative loss function of the form $\mathrm{Q}$=1.2${\mathrm{n}}^2_H\Lambda(\mathrm{T})$ above 10,000 K (optically thin plasma) \citep{2008ApJ...689..585C}. Below that value, we set $\Lambda(\mathrm{T})$ to be zero. Density, energy, momentum components (y and z), magnetic field (B$_y$ and B$_z$) are set as symmetric, while v$_x$ and B$_y$ are taken antisymmetric at the left and right boundaries.

To synthesize the observational features of SDO/AIA channel we use the FoMo code\footnote{\url{https://wiki.esat.kuleuven.be/FoMo/FrontPage}} to perform forward modelling \citep{2013A&A...555A..74A,2014ApJ...787L..22A,2015ApJ...806...81A,2015ApJ...807...98Y}. Using the AIA temperature response function \citep{2011A&A...535A..46D,2012SoPh..275...41B} the FoMo code converts the density to the intensity. We have synthesized the AIA 94~~\r{A} channel emission which has a characteristics log(T)~$\approx$6.8 .

\begin{figure*}[!htbp]
\centering
\includegraphics[width=1.0\textwidth]{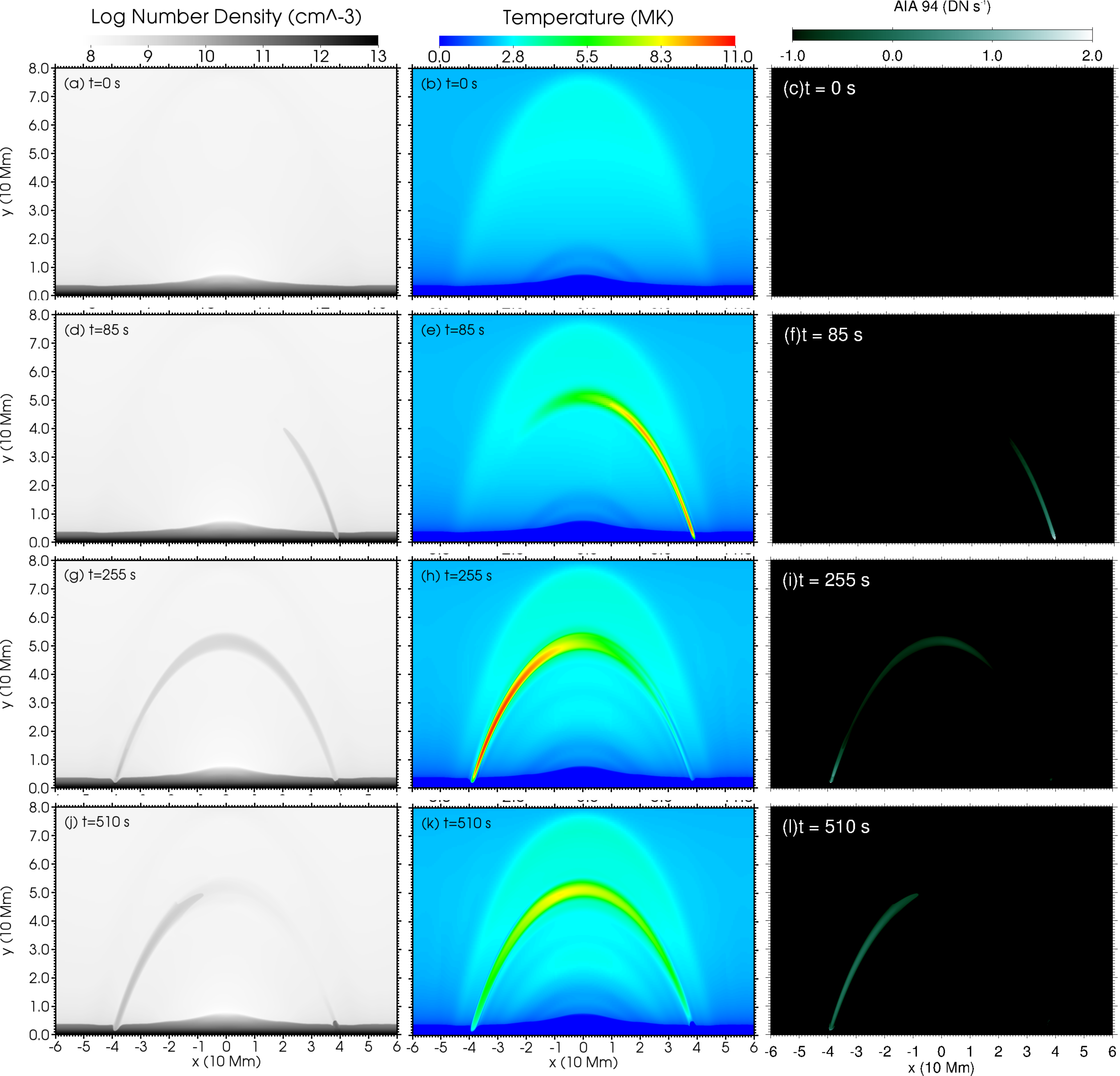}
\caption{ Snapshots showing the density ($n_e$), temperature (T) and the AIA 94~\r{A} intensity images respectively at different time of the simulation. The association of the intensity enhancement with the density and temperature is seen very clearly. An animated version (movie 7) is also available online.}
\label{numerical_context} 
\end{figure*}


\subsection{Analysis of the synthesized data}\label{sec4}

\begin{figure*}[!htbp]
 \centering
  \includegraphics[ angle=90,width=0.9\textwidth]{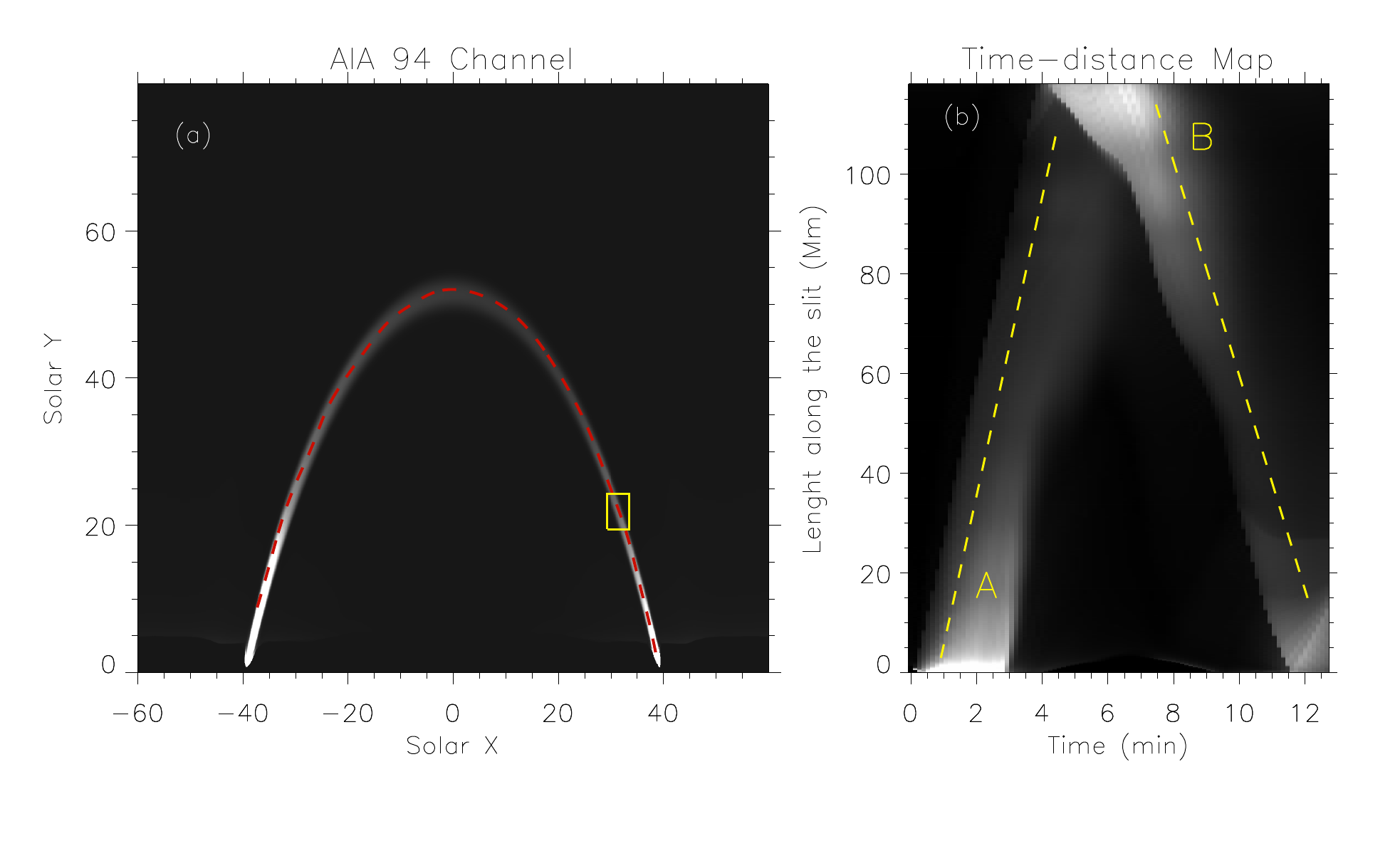}
 \caption{Panel (a) shows the time averaged synthesized AIA~94~\r{A} intensity image. The red dashed line indicates the artificial slit used for constructing the time-distance map. The yellow rectangular box is used to generate the intensity profile shown in Fig~\ref{simulation_damping}. Panel (b) shows the time-distance map with the fitted straight (in yellow) line used for speed calculation.}
 \label{f1}
\end{figure*}

 From the movie (movie~7, available online) we see that the wave propagates back and forth before fading out of the loop. Figure~\ref{numerical_context} shows the time evolution of the density,temperature and the AIA 94~\r{A} channel intensity. To see the wave propagation along the loop we put an artificial slit tracing the loop (red dashed line in the left panel of Fig.~\ref{f1}) to generate the time-distance map (right panel of Fig.~\ref{f1}). From the map we see clear signatures of reflection in the AIA 94~~\r{A} intensity images. The positions of the local maxima were identified
along each ridge and fitted with a linear function to calculate the propagation speed.
 
 The two yellow dashed lines represent the fitted straight line for the forward and the reflected wave having speeds 499 km s$^{-1}$ (line A) and 357 km s$^{-1}$(line B) ( with errors less than 15 km s$^{-1}$). These speeds compare very well with the average speeds we estimated from time-distance created using the XRT and AIA images. Here we must emphasize the fact that the simulation is 2D, so there is no projection effect unlike our observations. This consistency of the speed value validate our result obtained from DEM analysis which shows a temperature~$\approx$~10$\mathrm{MK}$.

\begin{figure*}[!htbp]
 \centering
  \includegraphics[ angle=90,width=0.7\textwidth]{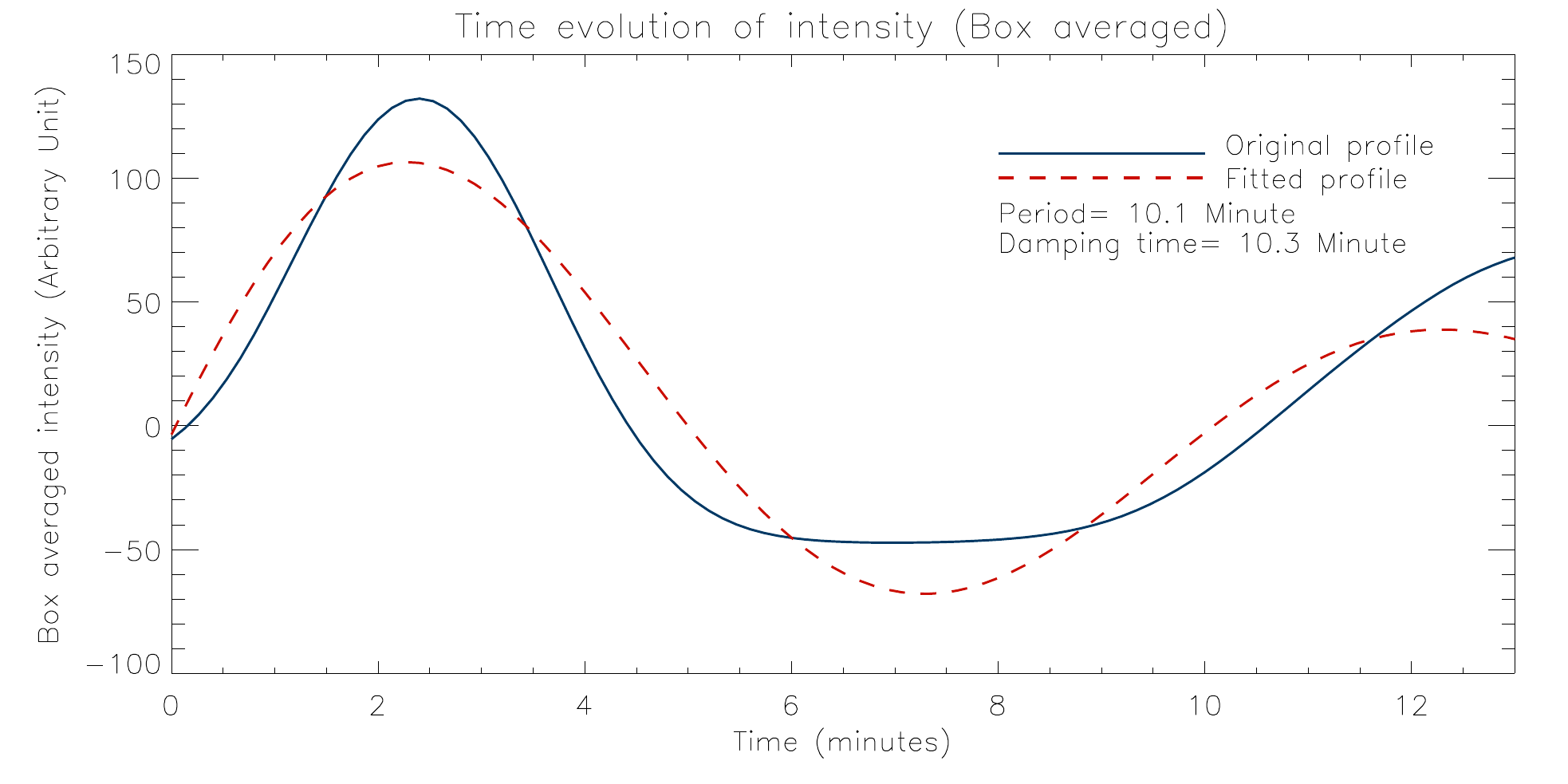}
 \caption{The blue solid line shows the intensity profile averaged over the chosen yellow box. The fitted damped sinusoidal function (Equation~\ref{damp_equ}) is shown with the red dashed line. The period and the damping time is printed on the panel.}
 \label{simulation_damping}
\end{figure*}

 To estimate the damping of the obsevred propagating intensity, we take a similar approach as for the observational data analysis. The time evolution of the intensity profile averaged over a chosen box (shown as yellow rectangle  in Fig~\ref{f1}) is plotted as blue solid curve in Figure~\ref{simulation_damping}. We fit the function (Equation~\ref{damp_equ}) on that profile and obtained the best fit curve, shown as the red dotted line. The period and the damping times are 10.1 minutes and 10.3 minutes respectively. These values are comparable with the values obtained for the event on 10$^{th}$ December. 

\section{Summary and conclusion}\label{sec5}

In this paper we report, for the first time, simultaneous observation of propagating and reflecting intensity disturbance in a hot coronal loop as seen with HINODE/XRT and SDO/AIA. We also report three other cases of such reflections of the propagating disturbances from XRT. Analysis of the events shows that the observed waves appear after a micro-flare occurs at one of the loop footpoints. The DEM analysis performed on the AIA image sequence revealed that the loop temperature and density to be 10~$\mathrm{MK}$ and $\approx$10$^9~$cm$^{-3}$ respectively. The average speed of propagation, as estimated from the time-distance maps, is lower than or comparable to the sound speed of the local medium estimated from the DEM analysis. This classifies the wave to be a slow propagating mode. We have studied in depth about the generation mechanism of such waves and also propose a scenario where a rapidly heated chromospheric plasma acts as a source for the high speed flows we observed in the coronal channels. 

 From the base-difference images and the time-distance maps we find that the waves are subject to a very fast damping as they propagate along the loop. We have quantitatively measured the damping and found out that the damping time ( 1/e fall from the initial amplitude) is almost equal to the period of the wave. 

We implemented these aspects in a numerical setup where the input loop parameters are the same as obtained from our observations. Injecting a small energy pulse at one of the loop footpoints which mimics the micro-flare trigger, we find a slow MHD wave propagating through the loop and gets reflected back from the other footpoint. We have introduced thermal conduction as the dominant damping factor and found that the wave actually damps very quickly as observed in the our events. Analyzing the synthesized data, we have obtained the propagation speeds which matches well with the observed speeds from XRT and AIA.  Using the observed density and temperature values we have reproduced similar time period of the propagating wave from our synthesized AIA data. We also find a good match between the damping time calculated from the synthesized data with the observed damping time. The wave seem to get damped quickly as it propagates through the loop (damping time comparable to the wave period). This fast damping could only be explained by the thermal conduction acting as the major damping mechanism for the propagating slow MHD mode along with the contribution from radiative cooling. \citet{2008ApJ...689..585C} have estimated the radiation loss from the coronal plasma using different abundance ratio and models and found that the loss due to radiation to be $\sim$10$^{-22}$ erg s$^{-1}$ cm$^{3}$. This loss is small compared to the total energy output from the flare which is $\sim$10$^{28}$ erg s$^{-1}$.

In conclusion we have identified and studied a set of unique events from HINODE/XRT (and SDO/AIA) where a micro-flare excites a slow MHD wave in a hot coronal loop and the wave gets reflected from other footpoint before fading away. Further study with simultaneous imaging and spectroscopic data where we can quantify the changes in the time evolution of the line patramets and their relation to the obsevred brightening will greatly improve our understanding about the origin as well as the propagation properties of such phenomena.

\section{Acknowledgment}
 SDO is a mission for NASA Living With a Star (LWS) program. Hinode is a Japanese mission developed and launched by ISAS/JAXA, with NAOJ as domestic partner and NASA and STFC (UK) as international partners. It is operated by these agencies in co-operation with ESA and the NSC (Norway). This research has been made possible by the topping-up grant CHARM+top-up COR-SEIS of the BELSPO and the Indian DST. It was also sponsored by an Odysseus grant of the FWO Vlaanderen, and Belspo's IUAP CHARM.
 \begin{table}[H]
\begin{center}
\centering
\caption{  Details of the XRT (`Be-thin') observations}  
\vspace{1cm}
\label{obs_details}
\begin{tabular}{lcccc r@{   }l c} 
  \hline
  & \multicolumn{1}{c}{Date}& \multicolumn{1}{c}{Time (UT)} & \multicolumn{1}{c}{Active Region} & \multicolumn{1}{c}{FOV} & \multicolumn{1}{c}{Cadence (s)} & \multicolumn{1}{c}{Pixel scale} \\
     
     \hline
     &  10-Dec-2015 & $ $ $ $ $ $ $ $ 04:30-05:16  $ $ $ $&  AR 12465 & 2106$''\times$2106$''$   &121 $ $ $ $ $ $ $ $ $ $ $ $& $ $ $ $ $ $ $ $ $ $ $ $ $ $4.1$''$\\
     &  22-Jan-2013 & $ $ $ $ $ $ $ $ 08:30-09:29  $ $ $ $&  AR 11654 & 394$''\times$394$''$   &61 $ $ $ $ $ $ $ $ $ $ $ $& $ $ $ $ $ $ $ $ $ $ $ $ $ $1.03$''$\\
     &  27-Jan-2013 & $ $ $ $ $ $ $ $ 20:09-20:59  $ $ $ $&  AR 11661 & 394$''\times$394$''$   &61 $ $ $ $ $ $ $ $ $ $ $ $& $ $ $ $ $ $ $ $ $ $ $ $ $ $1.03$''$\\
  \hline

\end{tabular}
\end{center}
\end{table}

 \bibliographystyle{apj}

\end{document}